\documentclass[12pt]{article}
\usepackage{graphicx}
\usepackage[margin=1.25in]{geometry}
\usepackage[usenames,dvipsnames]{color}
\usepackage{url}
\usepackage{caption,subcaption, cite}
\usepackage[colorlinks = true,
            linkcolor = blue,
            urlcolor  = blue,
            citecolor = blue,
            anchorcolor = blue]{hyperref}

\newcommand\pubnumber{ DESY-17-155\\ 
 KEK Preprint 2017-31\\
LAL 17-059\\
 SLAC--PUB--17161}
\newcommand\pubdate{October 2017}

\def\KEK{High Energy Accelerator Research Organization (KEK), Tsukuba,
  Ibaraki, JAPAN  }
\def\Tokyo{ICEPP, University of Tokyo, Hongo, Bunkyo-ku, Tokyo,
  113-0033, JAPAN}
\def\SNU{Dept. of Physics and Astronomy, Seoul National
  Univ.,  Seoul 08826, KOREA}
\def\DESY{DESY, Notkestrasse 85, 22607 Hamburg, GERMANY}
\def\Berlin{Institut f\"ur Physik, Humboldt-Universit\"at zu Berlin, 12489 Berlin, GERMANY}
\def\SLAC{SLAC,
    Stanford University, Menlo Park, CA 94025, USA}
\def\Tsinghua{Center for High Energy Physics, Tsinghua University,
  Beijing, CHINA}
\def\Osaka{Department of Physics, Osaka University, Machikaneyama, Toyonaka, Osaka 560-0043, JAPAN}
\def\IPMU{Kavli Institute for the Physics and Mathematics of the Universe,
University of Tokyo, Kashiwa 277-8583, JAPAN}
\def\Cornell{Laboratory for Elementary Particle Physics, Cornell
  University, Ithaca, NY 14853, USA}
\def\Orsay{LAL, Centre Scientifique d'Orsay, Universit\'e Paris-Sud, F-91898 Orsay CEDEX,
FRANCE}
\def\Munich{Max-Planck-Institut f\"ur Physik, F\"ohringer Ring 6,
  80805 Munich, GERMANY}
\def\Michigan{Michigan Center for Theoretical Physics, University of Michigan, Ann Arbor,
MI 48109, USA}
\def\UTA{Department of Physics, University of Texas, Arlington, TX
  76019, USA}
\def\Oregon{Center for High Energy Physics, University of Oregon, Eugene, Oregon
97403-1274, USA}
\def\Berkeley{ Department of Physics, University of California, Berkeley, CA 94720, USA}
\def\LBNL{Theoretical Physics Group, Lawrence Berkeley National Laboratory, Berkeley,
CA 94720, USA}

\def\Title#1{\begin{center} {\Large #1 } \end{center}}
\def\Author#1{\begin{center}{ \sc #1} \end{center}}

\newcommand\pubblock{\rightline{\begin{tabular}{l} \pubnumber\\
         \pubdate \end{tabular}}}
\newenvironment{Abstract}{\begin{quotation} \begin{center}
                       ABSTRACT
     \end{center}\bigskip  }{\end{quotation}}

\def\Acknowledgements{\bigskip  \bigskip \begin{center} \begin{large}
             \bf ACKNOWLEDGEMENTS \end{large}\end{center}}

\def\beq{\begin{equation}}
\def\eeq#1{\label{#1}\end{equation}}
\def\eeqn{\end{equation}}

\newenvironment{Eqnarray}%
   {\arraycolsep 0.14em\begin{eqnarray}}{\end{eqnarray}}
\def\beqa{\begin{Eqnarray}}
\def\eeqa#1{\label{#1}\end{Eqnarray}}
\def\eeqan{\end{Eqnarray}}
\def\CR{\nonumber \\ }

\def\leqn#1{(\ref{#1})}

\let\bar=\overbar

\def\etal{{\it et al.}}
\def\ie{{\it i.e.}}

\def\lsim{\mathrel{\raise.3ex\hbox{$<$\kern-.75em\lower1ex\hbox{$\sim$}}}}
\def\gsim{\mathrel{\raise.3ex\hbox{$>$\kern-.75em\lower1ex\hbox{$\sim$}}}}

\def\Im{{\rm Im}}
\def\Re{{\rm Re}}

\def\L{{\cal L}}

\def\L{{\cal L}}

\def\half{\frac{1}{2}}

\def\del{\partial}
\def\Dslash{\not{\hbox{\kern-4pt $D$}}}
\def\dslash{\not{\hbox{\kern-2pt $\del$}}}

\def\MET{\not{\hbox{\kern-4pt $E$}}_T}

\def\Dlr{\mathrel{\raise1.5ex\hbox{$\leftrightarrow$\kern-1em\lower1.5ex\hbox{$D$}}}}

\def\ee{e^+e^-}
\def\sstw{\sin^2\theta_w}

\def\msb{{\bar{\scriptsize M \kern -1pt S}}}

\def\drb{{\bar{\scriptsize D \kern -1pt R}}}
\def\ELER{e^-_Le^+_R}
\def\EREL{e^-_Re^+_L}

\def\eps{\epsilon}

\def\neu#1{\widetilde\chi^0_{#1}}

\makeatletter
\def\section{\@startsection{section}{0}{\z@}{5.5ex plus .5ex minus
 1.5ex}{2.3ex plus .2ex}{\large\bf}}
\def\subsection{\@startsection{subsection}{1}{\z@}{3.5ex plus .5ex minus
 1.5ex}{1.3ex plus .2ex}{\normalsize\bf}}
\def\subsubsection{\@startsection{subsubsection}{2}{\z@}{-3.5ex plus
-1ex minus  -.2ex}{2.3ex plus .2ex}{\normalsize\sl}}

\renewcommand{\@makecaption}[2]{%
   \vskip 10pt
   \setbox\@tempboxa\hbox{\small #1: #2}
   \ifdim \wd\@tempboxa >\hsize     
       \small #1: #2\par          
     \else                        
       \hbox to\hsize{\hfil\box\@tempboxa\hfil}
   \fi}

\makeatother

\def\met{\not{\hbox{\kern-4pt $E$}}_T}

\begin{document}
\begin{titlepage}
\pubblock

\vfill
\Title{Physics Case for the 250~GeV Stage\\  of the International Linear Collider}
\vfill
 \Author{LCC Physics Working Group}

\bigskip

\bigskip

\Author{Keisuke Fujii$^1$, Christophe Grojean$^{2,3}$, Michael
  E. Peskin$^4$  (Conveners); Tim Barklow$^4$, Yuanning Gao$^5$,
  Shinya Kanemura$^6$, Hyungdo Kim$^7$, Jenny List$^2$, Mihoko
  Nojiri$^{1,8}$, Maxim Perelstein$^{9}$, Roman P\"oschl$^{10}$,
  J\"urgen Reuter$^{2}$, Frank Simon$^{11}$, Tomohiko Tanabe$^{12}$,
  James D. Wells$^{13}$, Jaehoon Yu$^{14}$;  Mikael Berggren$^{2}$,\\
Moritz Habermehl$^{2}$,
  Sunghoon Jung$^{7}$, Robert Karl$^{2}$,\\ Tomohisa Ogawa$^{1}$,
  Junping Tian$^{12}$;
  James Brau$^{15}$,\\  Hitoshi Murayama$^{8,16,17}$ (ex officio)}

\vfill
\begin{Abstract}
The International Linear Collider is now proposed with a staged
machine design, with the first stage at 250~GeV with a luminosity
goal of 2~ab$^{-1}$.   In this paper, we review the physics
expectations for this machine.  These include precision measurements
of Higgs boson couplings, searches for exotic Higgs decays, other searches
for  particles that decay with zero or small visible energy, and
measurements of $\ee$ annihilation to $W^+W^-$ and 2-fermion states with improved
 sensitivity.  A summary table gives projections for the achievable
levels of precision  based on the latest full simulation studies.

\end{Abstract}
\vfill

\end{titlepage}

\noindent $^1$  \KEK\\
$^2$  \DESY\\
$^3$  \Berlin\\
$^4$  \SLAC \\ 
$^5$   \Tsinghua\\
$^6$  \Osaka\\ 
$^7$  \SNU \\ 
$^8$  \IPMU \\
$^9$  \Cornell \\
$^{10}$  \Orsay\\
$^{11}$  \Munich \\ 
$^{12}$   \Tokyo \\
$^{13}$  \Michigan\\ 
$^{14}$  \UTA \\ 
$^{15}$   \Oregon\\
$^{16}$  \Berkeley\\ 
$^{17}$   \LBNL\\

\newpage

\tableofcontents

\def\thefootnote{\fnsymbol{footnote}}
\setcounter{footnote}{0}

\newpage

\hbox to\hsize{\null}

\newpage

\section{Introduction}

The International Linear Collider (ILC) is a linear electron-positron
collider
planned for physics exploration and precision measurements in the
energy region of 200--500~GeV.  This report summarizes the
expectations for measurements of the Higgs boson and searches for
physics
beyond the Standard Model in the  program of this
accelerator at 250~GeV in the center of mass.

 The physics potential of the ILC 
is known to be very impressive.   A detailed accounting of the
expectations for this machine was presented in 2013 as a part of the
ILC  Technical Design Report~\cite{Behnke:2013xla,Baer:2013cma}
 and in white papers prepared
for the American Physical  Society's study of the future of US
particle physics 
(Snowmass 
2013)~\cite{SnowmassHiggs,Snowmasstop,SnowmassBSM,SnowmassEW}.  As
the ILC experiments have been studied in more detail, our Working
Group has published updated expectations for the general ILC
program~\cite{Fujii:2015jha} and for the direct search for new particles at
this collider~\cite{Fujii:2017ekh}.

In the past year, the program of the ILC has been reshaped in the
expectation of an international agreement and start of construction.
The Linear Collider Collaboration has recast the project as a staged
program with the first stage at 250~GeV~\cite{StagingRpt}.  This
 would significantly
lower the initial cost of the machine and provide a focused,
nearer-term goal for the project.  In this approach, the 250~GeV stage
of the ILC needs to be justified on its own merit rather than as a
part of a broader program that includes running at higher energies.
At the same time, new studies have revealed a very strong physics
potential for the 250~GeV stage of the ILC that was not specifically
emphasized in the reports cited above.  The purpose of this  article
is to summarize the case for the 250~GeV machine stage as it is
understood today.  We will see that there is a compelling
physics case  for the ILC that applies already at its 250~GeV stage.

Section 2 of this report updates the 2015 report on ILC operating
scenarios~\cite{Barklow:2015tja}, giving estimates of time vs. integrated
luminosity for an ILC project with 250~GeV, top quark threshold, and
500~GeV stages.

The most important objective of a 250~GeV $\ee$ collider is to make
precision measurements of the couplings of the 125~GeV Higgs boson to 
vector bosons, quarks, and leptons.  Unlike the situation at proton
colliders,  all of the major Standard Model decay modes of the Higgs
boson will be individually identifiable  in $\ee$ experiments.  This means that it
is possible to extract the absolute strengths of Higgs boson couplings
to high precision in a model-independent analysis.   In Sections 3 and 4
of this report, we will explain how this can be done and give
projected errors for the coupling constant determinations.

The search for new physics beyond the Standard Model is probably the
most important goal of particle physics today.   The LHC experiments
are carrying out intensive searches for new particle of many types,
and dark matter detection experiments add to the
variety of searches.   Because shifts of the
Higgs couplings can be induced by mixing with or loop corrections from
very heavy particles, the study of these couplings gives a route to new
physics that is essentially orthogonal to these methods. Today, the
LHC experiments are probing for large shifts of the Higgs couplings,
but, in typical models, the shifts of the Higgs couplings from their
Standard Model values are predicted to be small, at the 10\%
level and below. Thus, high-precision experiments, beyond the
expectations for LHC, are needed.   {\it  In our
opinion, this precision study of the Higgs boson is the most important 
suggested probe for new physics beyond the Standard Model that is 
not currently being exploited.}
This gives special impetus to the construction of a new accelerator
for precision Higgs studies.

 Qualitatively different models of new physics
predict different patterns of deviation from the Standard Model
prediction.  If the Higgs couplings can be measured individually
with high precision, it is possible to read the pattern and obtain
information on the properties of the new physics model.  We will
expand on  this point and present some examples in Section 5.

The Higgs boson provides another possible window into new physics.
Potentially, it is easy for the Higgs boson to couple to new particles
with no Standard Model interactions, particles that might make up the
dark matter or might otherwise be hidden from experiments that rely
on  other
probes.  We will review the ILC capabilities for the discovery of invisible
and exotic Higgs decays in Section 6.

In Sections 7, 8, and 9, we will discuss the capabilities of a 250~GeV $\ee$
collider beyond its program on the Higgs boson.  Section 7 will review
the
reach of such a machine for observation of the direct pair production
of dark matter particles and other particles difficult to detect at
the LHC.    In Section 8, we will discuss the new information that will be
available from the precision study of $\ee\to W^+W^-$.  In Section 9,
we will review the ability of $\ee$ annihilation to fermion-antifermion pairs
at 250~GeV to probe for new boson resonances and quark and lepton
substructure.   

Finally, in Section 10, we will review very briefly the
capabilities of the ILC, after an energy upgrade,
 for measurements at 350~GeV, 500~GeV, and
higher energies.  Indeed, the infrastructure of the ILC will support a
long future of 
experiments with $\ee$ collisions that would build on the success of
the first 250~GeV stage.

 An appendix gives a table of the projected measurement errors for the
 most important parameters. We recommend that these are the numbers 
that should be used in
 discussions of the ILC physics prospects and in
 comparisons of the ILC with other proposed facilities.

\section{Plan for ILC evolution and staging}

\begin{figure}
\begin{center}
\includegraphics[width=0.50\hsize]{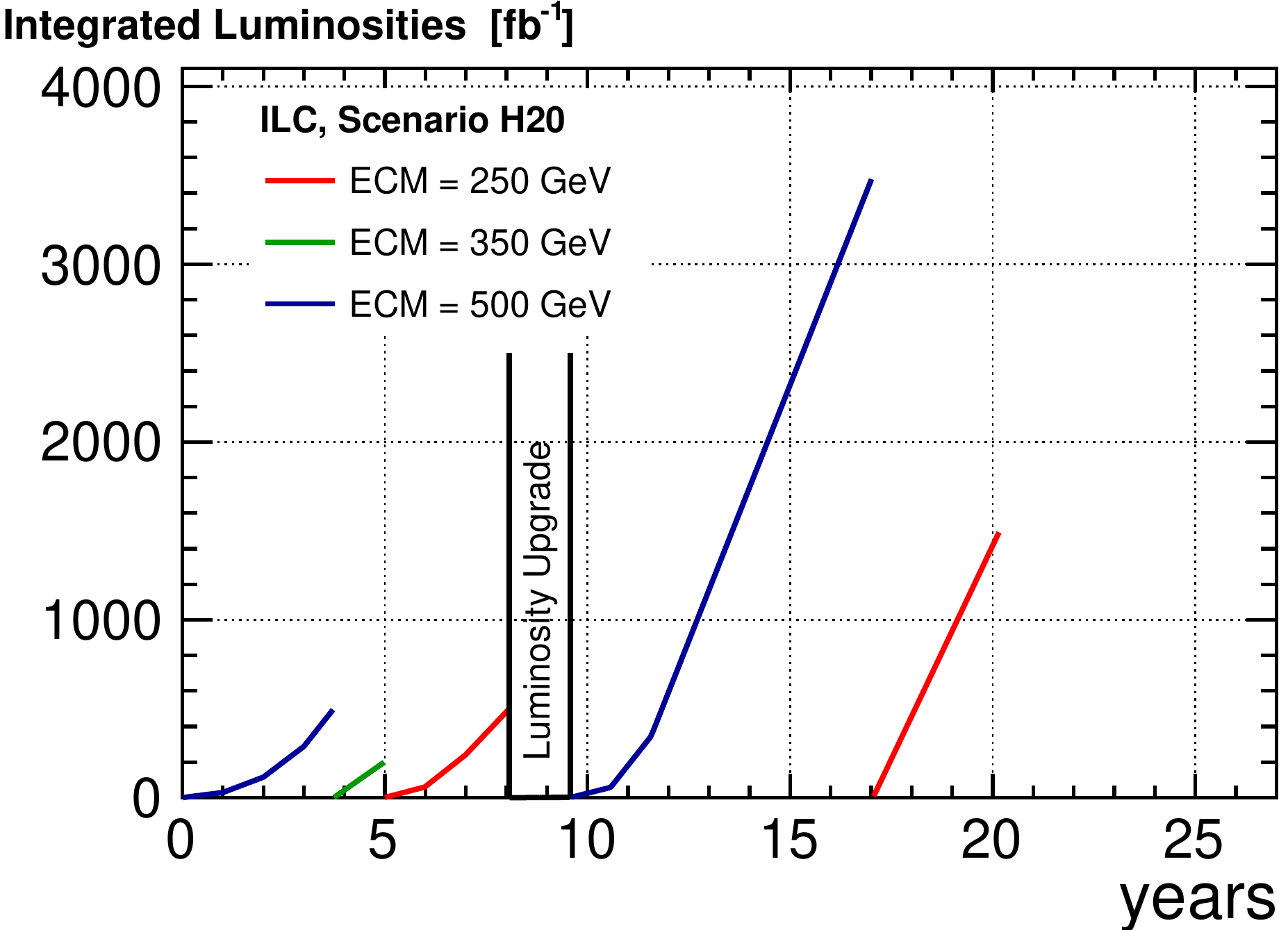}
\end{center}
\caption{The nominal 20-year running program
 for the 500-GeV ILC~\cite{Barklow:2015tja}.}
\label{fig:H20}
\end{figure}

Following the publication of the ILC Technical Design
Report~\cite{Behnke:2013xla,Baer:2013cma,Adolphsen:2013jya,Adolphsen:2013kya,Behnke:2013lya}, 
a canonical operating scenario
was defined for the ILC~\cite{Barklow:2015tja}. This operating
scenario assumed the construction of a 500-GeV 
machine, which within a 20-year period 
would accumulate integrated luminosities
of $4$\,ab$^{-1}$, $2$\,ab$^{-1}$ and $200$\,fb$^{-1}$ at
center-of-mass energies of $500$\,GeV, $250$\,GeV and $350$\,GeV, 
respectively, with beam polarizations of $\pm 80\%$
for the electron beam and  $\pm 30\%$ for the positron
beam. Figure~\ref{fig:H20} shows the time evolution 
of the data-taking envisioned in~\cite{Barklow:2015tja}, starting with operation at $500$\,GeV.
There were three main physics reasons for starting at $500$\,GeV:
first, the ability to use both of the major Higgs boson production
processes $\ee\to Zh$ and $\ee\to \nu\bar\nu h$ to measure Higgs
couplings;  second,  the ability to begin precision measurements of the
couplings of the 
top quark, including the direct measurement of the top-Yukawa coupling
from $t\bar t h$ production, and third, 
the ability to exploit the maximal discovery 
range for new particles.

Nevertheless, a very important part of the ILC physics program relies
on data collected in its running at 250~GeV, which already yields a
substantial sample of about half a million Higgs bosons tagged with recoiling
$Z$ bosons and subject to very small backgrounds.   Using new analyses
for reconstructing the various Higgs decay modes and a new, more
powerful theoretical approach, to be described in Section 3, we
realized that the 250~GeV program alone can already give powerful and
model-independent constraints on the Higgs properties.  Thus, a
staging scenario with a long first stage at 250~GeV makes sense from
the point of view of physics.  The purpose of this paper is to present
this argument in detail.

The detailed plan and accelerator design for the 250~GeV stage of the
ILC is described in~\cite{StagingRpt}.  In this section, we will
discuss the implications of this plan for the  running scenario and
luminosity expectations.

\begin{figure}
\begin{center}
   \begin{subfigure}{.49\hsize}
      \includegraphics[width=\textwidth]{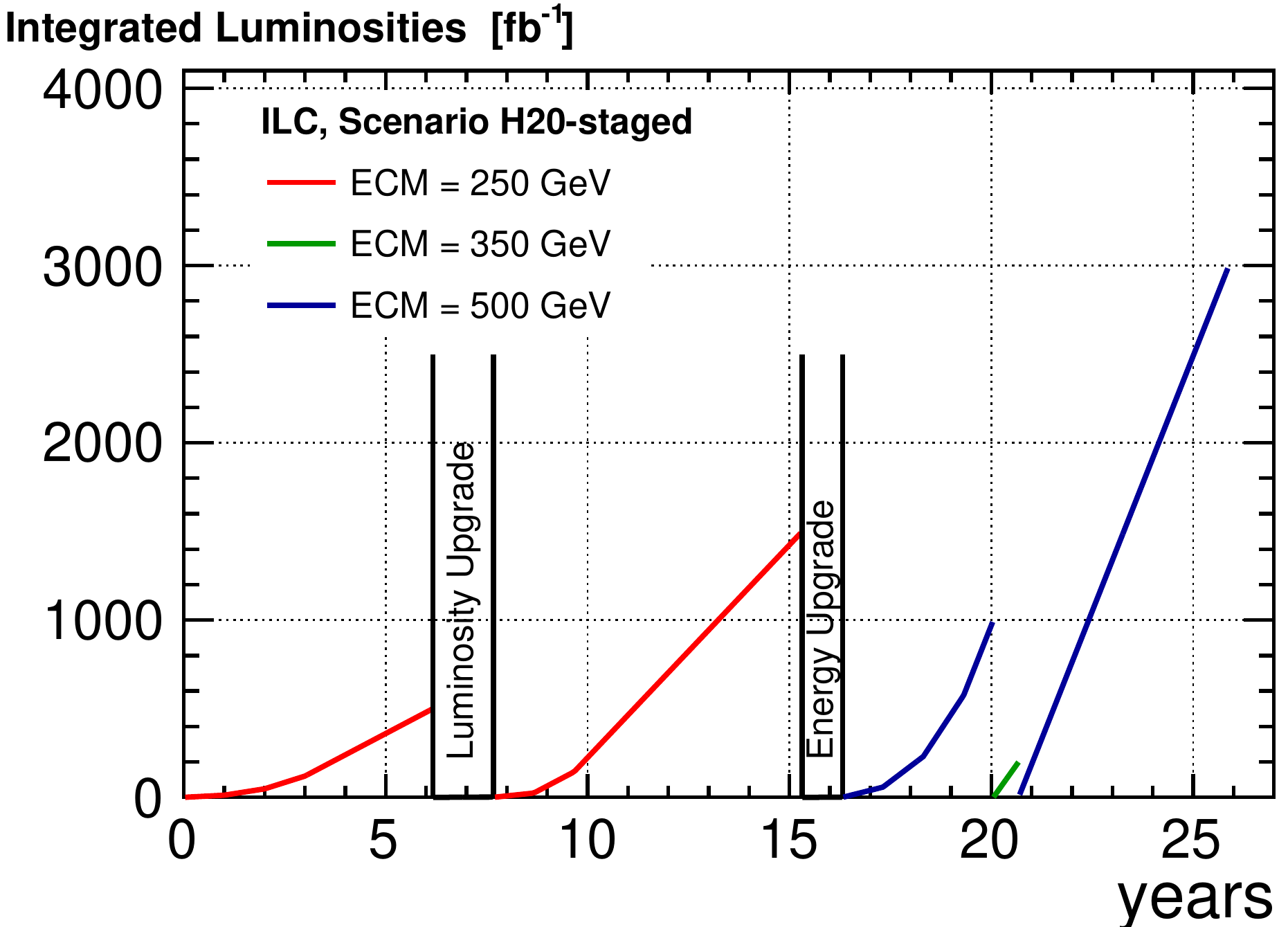}
         \subcaption{} \label{fig:H20stagedNOM}
   \end{subfigure} 
   \hspace{0.001\hsize} 
   \begin{subfigure}{.49\hsize}
      \includegraphics[width=\textwidth]{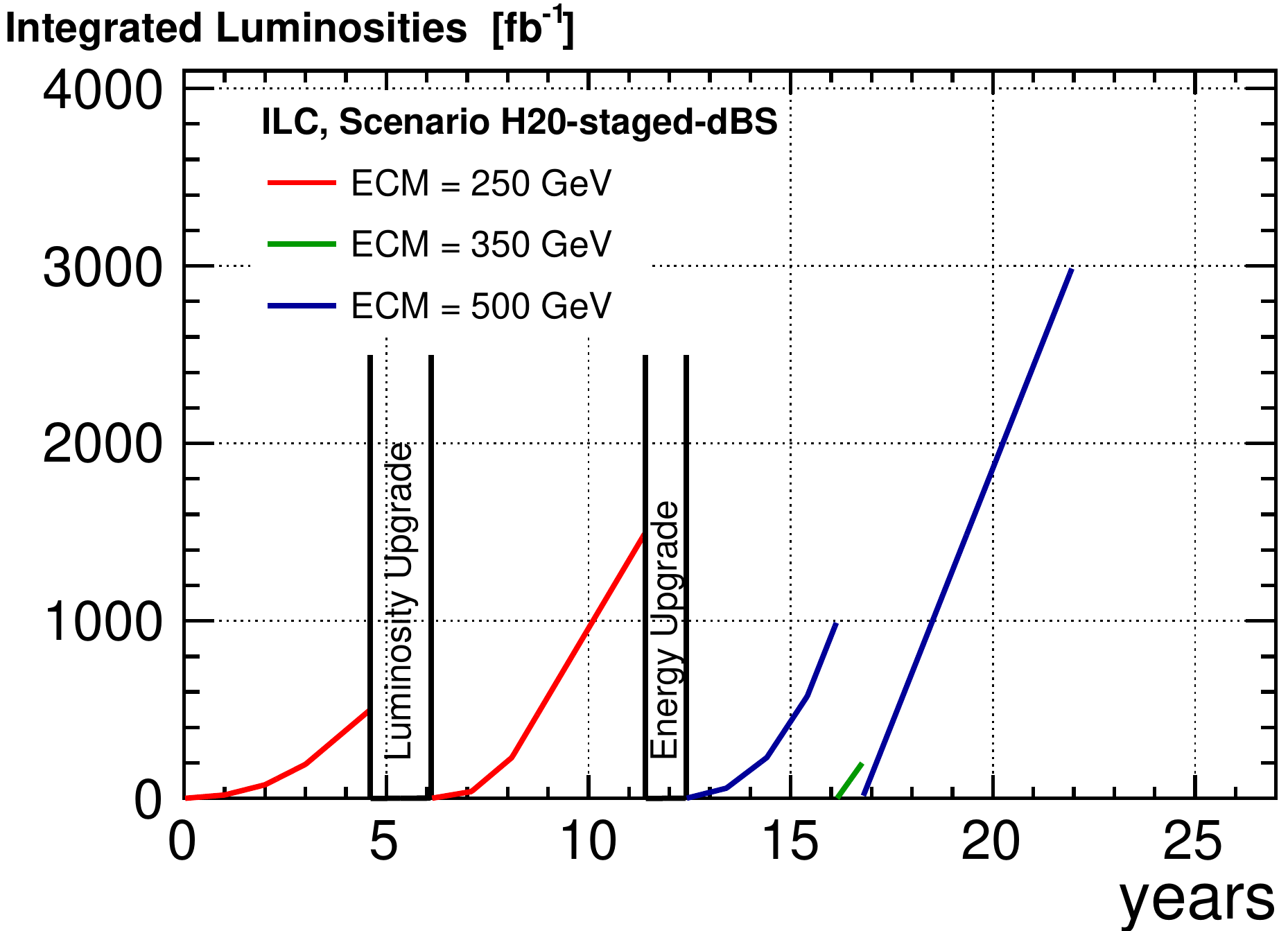}
         \subcaption{} \label{fig:H20stagedBS}
   \end{subfigure}  
\end{center}
\caption{Run plan for the staged ILC
 starting with a 250-GeV machine under two different
         assumptions on the achievable 
instantaneous luminosity at $250$\,GeV. Both cases 
         reach the same final integrated 
luminosities as in Fig.~\ref{fig:H20}.}
\label{fig:H20staged}
\end{figure}

Construction of a 250~GeV machine rather than the full 500~GeV machine
does change the expectation for the instantaneous luminosity that can
be assumed in 250~GeV running.   The original
running scenario (Fig.~\ref{fig:H20}) relied on the availability
of the full cryogenic and radio-frequency power of the $500$-GeV
machine in order to double the repetition rate from $5$ to $10$~Hz when
operating at $250$~GeV. This option is not available when
only half of  the power is installed in a minimal $250$~GeV
machine. Therefore the total operating time for accumulating the same
integrated luminosities as listed above stretches to 15 years of
operation for the 250~GeV stage and to 26 years for the full ILC
program.  This luminosity evolution is
shown in Fig.~\ref{fig:H20stagedNOM}.  

There is a cost-neutral
possibility
to increase the instantaneous luminosity 
by focussing the beam more strongly  at the IP.   This increases the 
level of beamstrahlung and $\ee$ pair production.  However, the ILC
interaction region is designed to cope with operation at 500~GeV and
even at 1~TeV.
Since  beamstrahlung is strongly
 energy-dependent, its effects at energies lower than these is 
much reduced, and so there is room for a more
aggressive choice of beam parameters at 250~GeV.  A revised set of 
accelerator parameters that implements this luminosity enhancement is
presented in Section 5 of \cite{StagingRpt}.
The effect of these new parameters on
the run plan is illustrated in Fig.~\ref{fig:H20stagedBS}. In this
plan, the length of the 250~GeV stage is 11 years and the 
total operating time for the full program is
 only slightly longer than the original 20
years. The exact effects of the new beam parameters on the detectors
and the physics measurements, taking account of the new beam energy
spectrum and  pair background,  still need to be evaluated 
quantitatively. All physics studies
quoted in this document 
are performed with the TDR parameters and thus apply strictly speaking to the
case shown in Fig.~\ref{fig:H20stagedNOM}. Nevertheless, 
we expect the differences to be 
small and are optimistic that similar results will 
be found with the new beam parameters 
corresponding to  Fig.~\ref{fig:H20stagedBS}.

It is well documented that beam polarization plays an essential role
in the physics program of the ILC at higher
energies~\cite{Baer:2013cma}.   The importance of having both electron
  and positron beam
  polarization at 250~GeV, for Higgs measurements and for other
  aspects of the ILC physics program, is discussed in some detail in
  \cite{Fujii:2018mli}.  Thus, in accord with the machine specifications
  presented in \cite{StagingRpt}, we assume  beam polarizations of 80\%
  and 30\% for the electron and positron beams of the 250~GeV ILC.
In~\cite{Barklow:2015tja}, 
the fractions of integrated
luminosity dedicated to each of the four possible sign combinations
were proposed for each center-of-mass energy. For operation at
$250$\,GeV,
 fractions of (67.5\%, 22.5\%, 5\%, 5\%)
were forseen for ($-+$,$+-$,$--$,$++$), where the first sign applies
to the electron beam polarization and the second to that of the
positron beam, giving emphasis to the $-+$ configuration as it has the
largest Higgs production cross section. In our new theory framework, the
left-right
 cross-section asymmetry plays an important role.  To optimize the
 measurement of this quantity, we assume in this paper
a sharing of (45\%, 45\%, 5\%, 5\%)  among the various beam
polarization choices.

\section{Effective Field Theory approach to 
precision measurements  at   $\ee$ colliders}

The goal of the ILC program on the Higgs boson is to provide
determinations of the various Higgs couplings that are both
high-precision and model-independent.  

 It is easy to see how this can
be achieved for some combinations of Higgs couplings.   In the
reaction $\ee\to Zh$, the Higgs boson is produced in association with
a $Z$ boson at a fixed lab-frame energy (110~GeV for $\sqrt{s}=
250$~GeV).  Up to small and calculable background from $\ee\to ZZ$ plus
radiation, observation of a $Z$ boson at this energy tags the presence
of a Higgs boson.  Then the total cross section for $\ee\to Zh$ can
be measured absolutely without reference to the Higgs boson decay
mode, and the various branching ratios of the Higgs boson can be
observed directly.

The difficulty comes when one wishes to obtain the absolute strength
of each Higgs coupling.   The coupling strength of the Higgs boson to
$A\bar A$ can be obtained from the partial width $\Gamma(h\to A\bar
A)$, which is  related to the branching ratio through
\beq
        BR(h\to A\bar
                 A) =      \Gamma(h\to A\bar A) /\Gamma_h \ , 
\eeqn
where $\Gamma_h$ is the total width of the Higgs boson.
In the Standard Model (SM), the width of a 125~GeV Higgs boson is 4.1~MeV,
a value too small to be measured directly from reaction kinematics.
So the width of the Higgs boson must be determined indirectly, and
this requires a model formalism.

In most of the literature on Higgs boson measurements at $\ee$
colliders, the width is determined using the $\kappa$ parametrization.
One assumes that the Higgs coupling to each species $A$ is modified from
the SM  value by a mutiplicative factor $\kappa_A$.  Then,
for example, 
\beq
      { \Gamma(h\to ZZ^*) \over SM} = \kappa_Z^2\ ,  \qquad   {
        \sigma(\ee\to Zh)\over SM} = \kappa_Z^2 \ . 
\eeq{kappavals}
where $SM$ denotes the SM prediction.  The $\ee$
environment offers a sufficient number of measurements to determine
all of the relevant parameters $\kappa_A$. In particular, the ratio
\beq
\sigma(\ee\to Zh)/ BR(h\to ZZ^*) 
\eeq{myratio}
is  independent of $\kappa_Z$ and directly yields the Higgs
width.  However, at the 250~GeV ILC even with 2 ab$^{-1}$ of data, the
statistics to measure $BR(h\to ZZ^*)$ is limited, and so the
precision of the width determination is compromised.   In the earlier
literature, including \cite{SnowmassHiggs,Fujii:2015jha}, this problem
was solved by using data from higher energies, making use of 
 the $W$ fusion reaction and the larger
and more precisely measurable branching ratio $BR(h\to WW^*)$. 

There is a more serious problem with the $\kappa$ formalism:   It is
not actually model-independent.   In principle, the Higgs boson can
have couplings to $ZZ$ with two different structures,
\beq
           \delta\L =     {m_Z^2\over v}(1 + \eta_Z)  h Z_\mu Z^\mu +
           \zeta_Z {1\over v} h Z_{\mu\nu} Z^{\mu\nu}  \ . 
\eeq{etazeta}
Here the coefficients $\eta_Z$, $\zeta_Z$ represent independent
corrections due to new physics effects.\footnote{In principle, additional
 structures can be formed by making $\eta_Z$ and $\zeta_Z$
  functions of momentum.  However, \leqn{etazeta} is the most general
  structure that appears in the SM perturbed by  dimension-6 operators
only, a restriction that we will make below.}  The Higgs boson coupling to
$WW$ has a similar structure, with parameters $\eta_W$, $\zeta_W$.
In the $\kappa$ formalism, the couplings $\zeta_Z, \zeta_W$ are 
 assumed to be zero.
The operator multiplying $\zeta_Z$ is 
 momentum-dependent, so the effect of this
term depends on the momentum configuration of the vector bosons.
Indeed, for a 125~GeV Higgs boson and $\sqrt{s}= 250$~GeV,
\beqa
          \Gamma(h\to ZZ^*)/SM &=& (1 + 2\eta_Z - 0.50 \zeta_Z) \CR
         \sigma(\ee\to Zh)/SM &=& (1 + 2\eta_Z + 5.7  \zeta_Z) \  . 
\eeqan
Then the $Z$ coupling information does not cancel out of \leqn{myratio}
and so this ratio does not determine the Higgs width unambiguously.

There is an attractive solution to this problem.  The fact that the
LHC experiments have not yet observed new particles due to physics
beyond the SM suggests that these particles are heavy,
with masses above 500 GeV for electroweakly coupled states and above 1
TeV for strongly interacting states.  If indeed new particles are
sufficient heavy, we can describe
the physics of the 125~GeV Higgs boson by integrating these particles
out of the Lagrangian
and replacing their effects by an expansion in operators
built of Standard Model fields.   The SM itself is the
most general gauge-invariant Lagrangian built of SM fields with
operators of dimension up to 4.  Corrections to the SM are then
described by the addition of operators of dimension 6 and higher.  If
the minimum mass of the new particles is $M$, 
operators of dimension 6 will have coefficients proportional to
$m_h^2/M^2$.  These represent the first order in an expansion in
$m_h^2/M^2$.  Possible operators of dimension 8 and higher are
multiplied by additional factors of $m_h^2/M^2$.  It is then suggested
to parametrize the effects of the most general new physics on the
Higgs boson by writing an effective Lagrangian that consists of the SM
Lagrangian plus the most general set of $SU(3)\times SU(2)\times
U(1)$-invariant dimension-6 operators.  This is called  the Standard Model
Effective Field Theory (EFT) formalism.

The EFT formalism has been accepted by the LHC community as the best
way to parametrize  deviations from the SM in Higgs physics and in
vector boson interactions
 that might be observed at
the LHC~\cite{deFlorian:2016spz}.  The advantage of this approach
for the LHC experiments is that it provides a precise theoretical
formalism in which radiative corrections can be computed.  This is
important at the LHC, because Higgs signatures often require suppressed
decay modes with contributions from different basic couplings (for
example, dileptons in the final state), and because quantitative
predictions for QCD processes require NLO corrections. However, it is
difficult to use this formalism in a completely general way at the
LHC.   The most important difficulty is that the the number
of possible dimension 6 operators is very large.   There are 59
dimension-6 operators that can be added to the SM Lagrangian even if
we restrict ourselves to one generation of fermions and to baryon
number-conserving operators. Most of these involve quark and
gluon
 fields
and are relevant to LHC reactions. 

For reactions that involve only SM vector bosons, Higgs bosons, and
light leptons, the number of possible operators is much smaller,
though still sizable.  In \cite{Barklow:2017awn}, it is argued that
the most general effects of high-mass new physics on these reactions can be
parametrized by 10 dimension-6 operators.\footnote{Of these 10
  operators, 1 shifts the triple- and quadruple-Higgs couplings but
  does not affect single-Higgs processes at the tree level.} 
The same 10 operators
parametrize the new physics contributions to precision electroweak
observables and to observables in $\ee\to W^+W^-$.  There are
sufficient measurements available to an $\ee$ collider to determine all
10 parameters without significant degeneracies.  This gives a unified
formalism for testing the SM, one that brings together the full set of
measurements available at an $\ee$ collider.   Inclusion of on-shell Higgs
decays brings in 7 additional operators.  Measurement of Higgs decays
allows the coefficients of these additional operators to be
determined also.   Fits to prospective $\ee$ collider data using the
EFT formalism have been  presented
 in \cite{Ge:2016zro,Ellis:2017kfi,Durieux:2017rsg,Barklow:2017suo}.
 The last of these papers emphasizes the completeness of the
 17-parameter model and the ability to fit the 17 parameters
 simultaneously using the expected data set from $\ee$ colliders.

An illustration of the power of this formalism is given by the answer
to the question posed at the beginning of this section. The problem,
again, is that the $hZZ$ and $hWW$ couplings each involve two separate
kinematic structures whose coefficients must be separately
determined.  The EFT
formalism contains coefficients of dimension-6 operators that
contribute to the $\eta_{Z,W}$ and $\zeta_{Z,W}$ parameters
 defined in \leqn{etazeta}.
However, the $SU(2)\times U(1)$-invariance of the EFT Lagrangian leads
to relations between the coefficients for $Z$ and $W$.   These
relations are not simple, but they turn out to be very constraining.
For the $\eta$ parameters,
\beqa
      \eta_W &=&  -\half c_H + 2 \delta m_W - \delta v \CR
     \eta_Z  &=&   -\half c_H + 2 \delta m_Z - \delta v  - c_T \ , 
\eeqan
where the $c_i$ are coefficients of dimension-6 operators, $\delta
m_W$, $\delta v$, and $\delta m_Z$ are combinations of these
coefficients that shift the parameters $m_W$, $G_F$, and $m_Z$ (and
are constrained  by the measured values of those quantities), and $c_T $
is essentially the $T$ parameter of 
precision electroweak formalism~\cite{Peskin:1990zt}
and is constrained to be small by precision electroweak measurements.
Similarly, 
\beqa
      \zeta_W &=& (8 c_{WW})\CR
     \zeta_Z  &=&  \cos^2\theta_w (8 c_{WW}) + 2 \sin^2\theta_w (8
     c_{WB}) +  (\sin^4\theta_w/\cos^2\theta_w)(8 c_{BB}) , 
\eeqan
in which the parameters $c_{WB}$, $c_{BB}$ also contribute to $\ee\to
W^+W^-$ and the Higgs decays to $\gamma\gamma$ and $Z\gamma$ and so
can be strongly constrained.   The network of constraints essentially reduces the
problem to be solved by ILC Higgs measurements to the determination of
the two parameters $c_H$, $c_{WW}$ using measurements of the process
$\ee\to Zh$ and the decay
$h\to WW^*$.  For both reactions, there will be  ample statistics at
the 250~GeV ILC.    A particular feature of interest is that new Higgs
observables not previously considered in ILC studies become relevant. 
 In particular,
the polarization asymmetry and angular distributions in $\ee\to Zh$
turn out to put  very strong constraints on $\zeta_Z$ or $c_{WW}$~\cite{Barklow:2017suo}.
 
Remarkably, then, the EFT formalism, applied to  the $\ee$ world, realizes in a
very beautiful way the hopes put forward by its proponents in the LHC
world.  It provides a single formalism that knits together constraints
from precision electroweak measurements and from all of the processes,
not only Higgs processes,
that are measured in high-energy $\ee$ reactions.   The number of free
parameters, describing the most general new physics perturbation, is
large but manageable, and all relevant parameters can be determined
independently.   Although second-order electroweak corrections to
already small perturbations are not obviously relevant, the  formalism also
provides a Lagrangian setting in which radiative corrections can be
computed unambiguously.  This formalism thus provides a powerful
method for stringent tests of the SM and, we hope, discovery of new,
beyond-SM effects.

\section{Measurement of Higgs boson couplings}

In the SM, all of the Higgs boson couplings are predicted in terms of the 
value of the mass of 
Higgs boson, which is now
known to 0.2\% accuracy at the LHC \cite{Aad:2015zhl}. The observation
of any 
deviation from these predictions 
would imply new physics beyond the Standard Model.  As was already
noted, and will be discussed further in Section~5, the expectations
for deviations are small in typical 
BSM scenarios.   It is thus one of the main goals
 for a future $e^+e^-$ collider is to achieve 
O(1\%) precision in the measurement of Higgs boson couplings. 
This goal has been demonstrated to be achievable at the
 ILC~\cite{Baer:2013cma,SnowmassHiggs,Fujii:2015jha,Barklow:2015tja}
for the running scenarios with a baseline of $\sqrt{s}=500$ GeV,
based on full detector simulations for most of the observables.

Thus we focus here on the  prospects for the 
measurement of Higgs boson couplings 
at the 250 GeV stage of ILC assuming a 
total integrated luminosity of 2 ab$^{-1}$.

In this section we will first introduce the basic observables that are
used to fit for Higgs boson couplings.   We will also discuss the expected
precisions of branching ratios, which can be determined free of any
theory assumptions.  To quote absolutely normalized couplings, we need
to determine the Higgs boson width, and this requires a theory
framework.  We will  discuss  the width determination in the $\kappa$ and
EFT formalisms described  in the previous section, emphasizing 
the major consequences of the change from  $\sqrt{s}=500$~GeV to 250~GeV
and the new observables that play an important role in  the EFT approach.
Unless explicitly stated, all numbers shown in this section are for a
total
 integrated luminosity of 2 ab$^{-1}$
and for beam polarization sharing as introduced in Section 2.

\begin{figure}
\begin{center}
\includegraphics[width=0.70\hsize]{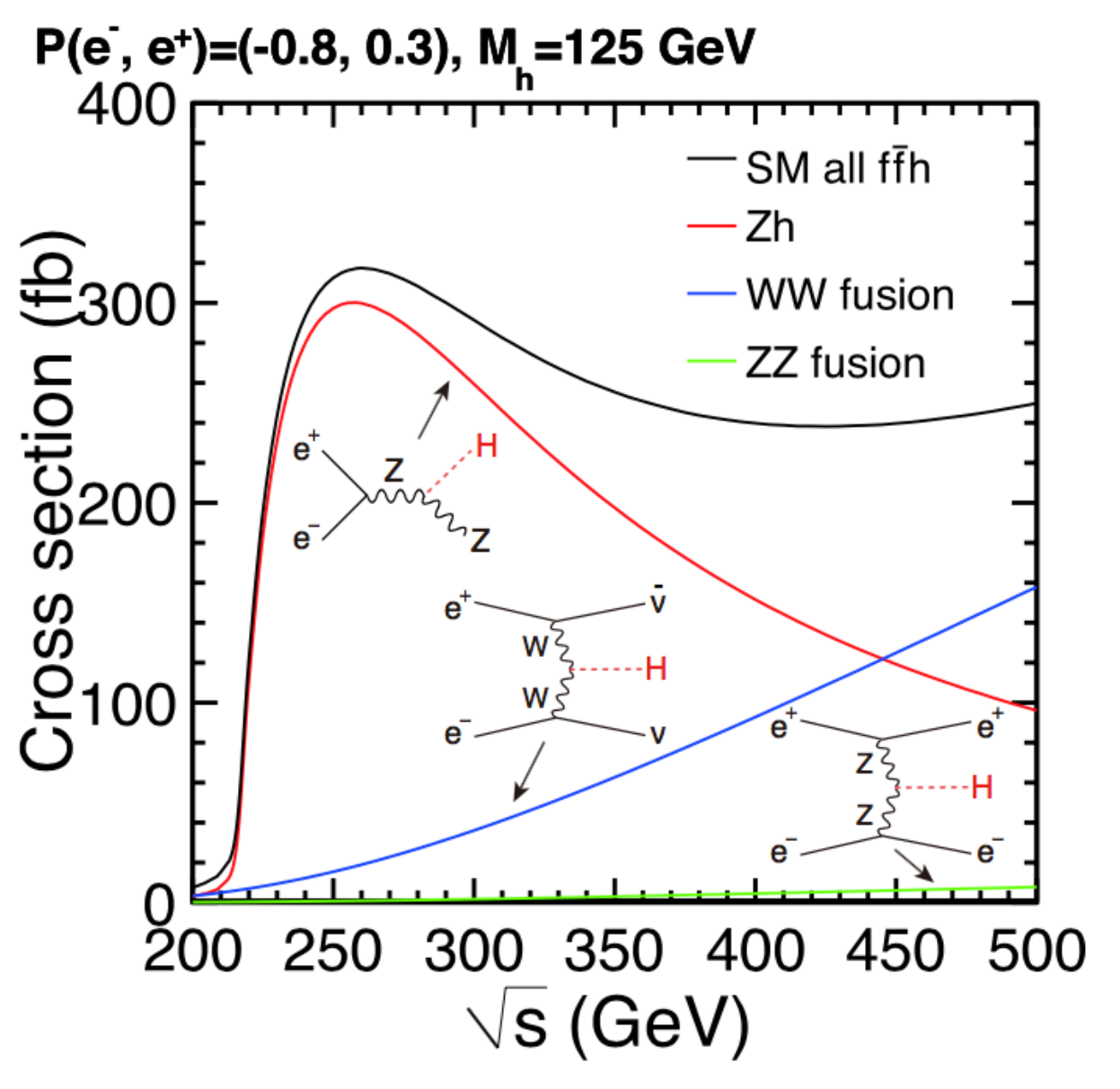}
\end{center}
\caption{Cross sections for the three major Higgs production processes
  as a function of 
center of mass energy, 
from~\cite{Baer:2013cma}.}
\label{fig:HiggsProdILC}
\end{figure}

\subsection{Basic observables: $\sigma$, $\sigma\cdot{BR}$}

\begin{figure}
\begin{center}
\includegraphics[width=0.95\hsize]{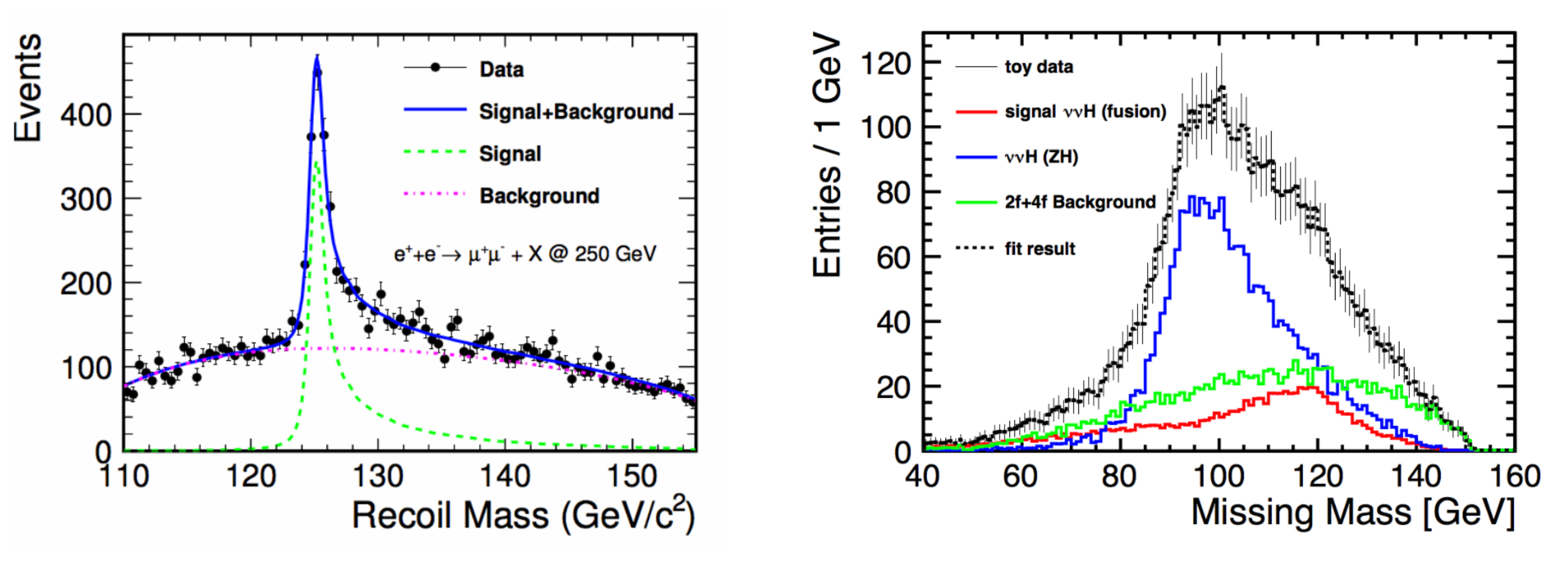}
\end{center}
  \caption{(left) recoil mass spectrum against
 $Z\to\mu^+\mu^-$ for signal $e^+e^-\to Zh$ and SM background 
  at 250 GeV \cite{Yan:2016xyx}; 
  (right) missing mass spectrum for the signal 
$e^+e^-\to\nu\bar\nu h, h\to b \bar{b}$ and the SM background 
  at 250 GeV \cite{H2bb1,H2bb2}.}
  \label{fig:Sigmah}
\end{figure}

The SM cross sections for the leading Higgs production processes
 in $\ee$ annihilation with $(P_e,P_p) = (-0.8, +0.3)$ polarized beams
are shown in Fig.~\ref{fig:HiggsProdILC}.   The process $\ee\to Zh$ 
attains its maximum cross section at $\sqrt{s}= 250$~GeV, 
providing about half a million $Zh$ events from an integrated luminosity of 2 ab$^{-1}$.
This allows the precise measurement of the inclusive cross section
$\sigma_{Zh}$, using 
the recoil mass technique,   and of  the rates $\sigma_{Zh}\cdot{BR}$
 for various decay modes.
Up-to-date estimates for measurements of $\sigma_{Zh}$ and 
$\sigma_{Zh}\cdot BR$
are given in the Appendix of \cite{Barklow:2017suo}. 
Most notably, $\sigma_{Zh}$ is measured to 1.0\% for both $(-+)$ and $(+-)$
initial polarization states
at $\sqrt{s}=250$ GeV.
An example of the recoil mass distribution in the $Z\to\mu^+\mu^-$ channel 
is given in Figure \ref{fig:Sigmah} (left).

With both $\sigma_{Zh}$ and $\sigma_{Zh}\cdot{BR}$ measured, 
the absolute branching ratios can be determined independently of any
fitting formula.  Among the SM branching ratios, the best measured ones
would be ${BR}_{bb}$ and ${BR}_{\tau\tau}$, with accuracies of 0.89\% and 1.4\% respectively. 
If there are O(1\%) or larger exotic decay modes, a first hint would already be provided 
by observing the resulting deviations in $BR_{bb}$ and $BR_{\tau\tau}$. 
The branching ratios 
${BR}_{cc}$ and ${BR}_{gg}$, which are very 
challenging to access directly at the LHC, 
can be measured to 3.2\% and 2.7\% respectively.
 ${BR}_{WW}$ and $BR_{ZZ}$, which play a special role in the total width determination,
can be measured to 1.9\% and 6.7\% respectively. 
The branching ratios to the rare decay modes, $BR_{\gamma\gamma}$ and
$BR_{\mu\mu}$ are limited by available statistics and  can be measured
only to
13\% and 27\% respectively.\footnote{A promising
 improvement to the $BR_{\mu\mu}$ estimate
is  presented in \cite{H2mumu}.}
However, these measurements can be improved by combination with LHC
results, since  the ratios of branching ratios
$BR_{ZZ}/BR_{\gamma\gamma}$ and $BR_{\mu\mu}/BR_{\gamma\gamma}$ are
expected to be measured at the HL-LHC, 
with accuracies of 2\% and 12\% \cite{H2aaLHC,H2mumuLHC},
respectively. 
The fact that $h$ is produced in recoil against a $Z$ boson gives
sensitivity to invisible decay modes of the Higgs boson sufficient to
provide a limit 
$BR_{inv}<0.32\%$ at the 95\%  C.L.  The sensitivity of the 250~GeV
program to invisible and exotic Higgs decays will be  discussed further in
Section 6. 

For the  $\sigma_{Zh}$ and $BR$ measurements, there seems to be 
no problem
in  going from 
$\sqrt{s}=500$ GeV to 250 GeV, despite the lower expected luminosity.
In fact $\sigma_{Zh}$ turns out to be better measured at 250 GeV,
mainly thanks to the larger cross section and
 less significant beamstrahlung effect.
On the other hand, the lowered energy 
is expected to have a significant impact  on the measurement of the  $WW$
fusion process ($\ee \to \nu\bar\nu h$),
the cross section of which becomes  almost a factor of 10 smaller. 
Moreover, due to the limited available phase space at 250 GeV, 
the missing mass spectrum in the $\nu\bar\nu h$ process is significantly
overlapping with that in the
$Zh, Z\to\nu\bar\nu$ process, as shown in Figure~\ref{fig:Sigmah} (right). 
As a result, $\sigma_{\nu\nu h}\cdot{BR}_{bb}$ for the $(-+)$
polarization state can 
only be
 measured to 4.3\%.  There is a correlation of $-34\%$
with the $\sigma_{Zh}\cdot{BR}_{bb}$ measurement, which is needed to
determine $BR_{bb}$.  This has only a tiny effect on the final result.

\subsection{Expected precisions for Higgs boson couplings in the
  $\kappa$ formalism}

Using only the basic observables introduced above, 
all of the Higgs boson couplings can be extracted via a global fit in the 
$\kappa$ formalism defined above \leqn{kappavals}.   The total width
of
the Higgs boson is given by 
\beq
\Gamma_h=\frac{\Gamma_{ZZ}}{BR_{ZZ}}=\frac{\Gamma_{WW}}{BR_{WW}},
\eeq{eqn:TotalWidth} 
where $\Gamma_{ZZ}$ ($\Gamma_{WW}$) is the partial decay width to $ZZ^*$ ($WW^*$).
In the $\kappa$ formalism, $\Gamma_{ZZ}$ ($\Gamma_{WW}$) is determined 
via $\kappa_Z$ ($\kappa_W$) from the measurement of $\sigma_{Zh}$ ($\sigma_{\nu\nu h}$) 
based on a simple relation,
\beq
\Gamma_{ZZ}\propto\kappa_Z^2\propto\sigma_{Zh}
~~(\Gamma_{WW}\propto\kappa_W^2\propto\sigma_{\nu\nu h}).
\eeq{eqn:PartialWidthWZ}
All the other couplings ($\kappa_A$) or partial decay widths
 ($\Gamma_{AA}$), e.g. $A=b,c,g,\tau,\mu,\gamma$,
are then determined as
\beq
\kappa_A^2\propto\Gamma_{AA}=\Gamma_h\cdot BR_{AA}.
\eeq{eqn:PartialWidthA}
As seen above, $BR_{ZZ}$ is only measured to $6.7\%$, so if only the first half of \leqn{eqn:TotalWidth}
is used, all Higgs boson couplings (except $\kappa_Z$) would have
an uncertainty greater than 3\%.
$BR_{WW}$ is 10 times larger than $BR_{ZZ}$ and so  can be measured
much more precisely.  For this reason, it is 
well recognized that in the $\kappa$ formalism 
the measurement of  the $WW$ fusion cross section $\sigma_{\nu\nu h}$
along with $BR_{WW}$  (using the second half of \leqn{eqn:TotalWidth})
is crucial for measurement of $\Gamma_h$ and of  all 
$\kappa_A$  with $A\neq Z$.
The expected precisions for Higgs boson 
couplings in the $\kappa$ formalism are
given in Table \ref{tab:higgscouplings}.  We see that, 
at $\sqrt{s}=250$ GeV, $\kappa_Z$ is determined very precisely, with
accuracy of 0.38\%,  but
most other $\kappa_A$ are determined 
 to no better than $\sim 2\%$ (limited by $\sigma_{\nu\nu h}$ 
and $BR_{ZZ}$ measurements).  An exception is $\kappa_\gamma$, which is helped 
significantly by the fact that the fit makes use of the expected
measurement of $BR_{ZZ}/BR_{\gamma\gamma}$ at the HL-LHC.

\begin{table}
\begin{center}
\begin{tabular} {lcccccc}
 & ILC250   &  & +ILC500  &  & \\
 & $\kappa$ fit   & EFT fit    & $\kappa$ fit    &  EFT fit \\
\hline
$g(hbb)$                           &    1.8        &  1.1         & 0.60          & 0.58 \\
$g(hcc)$                            &    2.4        &  1.9         & 1.2          & 1.2 \\
$g(hgg)$                           &    2.2        &  1.7         & 0.97          & 0.95 \\
$g(hWW)$                        &    1.8        &  0.67       & 0.40          & 0.34 \\
$g(h\tau\tau)$                   &    1.9        &  1.2         & 0.80          & 0.74 \\
$g(hZZ)$                          &    0.38       &  0.68       & 0.30          & 0.35 \\
$g(h\gamma\gamma)$     &    1.1        &  1.2         & 1.0            & 1.0 \\
$g(h\mu\mu)$                   &    5.6        &  5.6         & 5.1            & 5.1 \\
$g(h\gamma Z)$               &    16         &  6.6         & 16             & 2.6 \\
\hline
$g(hbb)/g(hWW)$             &    0.88      &  0.86         & 0.47        & 0.46 \\
$g(h\tau\tau)/g(hWW)$     &    1.0        &  1.0           & 0.65        & 0.65 \\
$g(hWW)/g(hZZ)$            &     1.7        &  0.07         & 0.26        & 0.05 \\
\hline
$\Gamma_h$                   &    3.9         &  2.5         & 1.7            & 1.6 \\
\hline
$BR(h\to inv)$                  &    0.32        &  0.32       & 0.29          & 0.29 \\
$BR(h\to other)$              &    1.6          &  1.6         & 1.3            & 1.2 \\
\end{tabular}
\caption{Projected relative errors for Higgs boson couplings and other Higgs observables, in \%,
for fits in the $\kappa$ and EFT formalisms.
 The ILC250 columns assume a total 
integrated luminosity of 2 ab$^{-1}$ at $\sqrt{s}=250$ GeV, shared by $(-+,+-,--,++)=(45\%,45\%,5\%,5\%)$ 
as described in Section~2. The ILC500 columns assume,
 in addition, a total integrated luminosity 
of 200 fb$^{-1}$ at $\sqrt{s}=350$ GeV, shared
 as $(45\%,45\%,5\%,5\%)$, and a total integrated luminosity 
of 4 ab$^{-1}$ at $\sqrt{s}=500$ GeV, shared as
$(40\%,40\%,10\%,10\%)$. Three observables at the HL-LHC,
$BR_{\gamma\gamma}/BR_{ZZ}$, 
$BR_{\gamma Z}/BR_{\gamma\gamma}$ and $BR_{\mu\mu}/BR_{\gamma\gamma}$,
are included in all of  the fits. The effective couplings 
$g(hWW)$ and $g(hZZ)$ are defined as proportional to the 
square root of the corresponding partial widths.
The last two lines give 95\% confidence upper limits on
 the exotic branching ratios.  The detailed formulae used in the EFT
 fit, and the resulting covariance matrix, can be found 
in \cite{Barklow:2017awn}.}
\label{tab:higgscouplings}
\end{center}
\end{table}

\subsection{Expected precisions for Higgs boson
 couplings in the EFT formalism}

In the EFT formalism, Higgs-$Z$ interaction consists of two distinct
Lorentz structures, shown in \leqn{etazeta}.   As explained in the
previous section,  \leqn{eqn:PartialWidthWZ} is violated by the
$\zeta_Z$ terms.  Thus, the $\kappa$ formalism is not
model-independent, and it is not as general as the EFT formalism.

\begin{figure}
\begin{center}
\includegraphics[width=0.99\hsize]{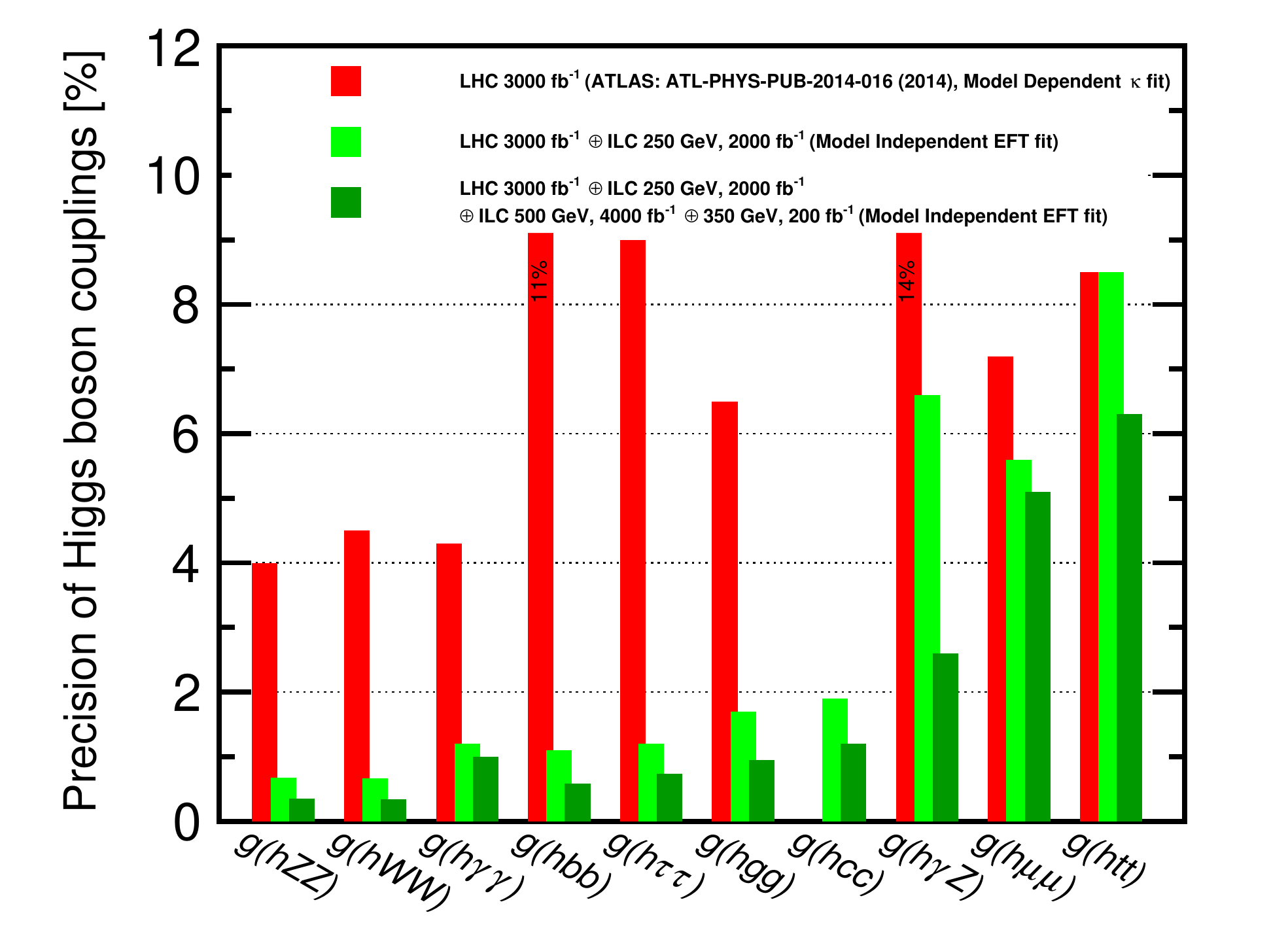}
\end{center}
\caption{Illustration of the Higgs boson coupling uncertainties
  from fits in the EFT formalism, as presented in
  Table~\ref{tab:higgscouplings}, and comparison of these projections to
  the results of model-dependent estimates for HL-LHC uncertainties 
presented by the ATLAS 
collaboration~\cite{H2aaLHC}.   Earlier projections for HL-LHC are
summarized in \cite{Dawson:2013bba}.}
\label{fig:Higgssummary}
\end{figure}

However, the EFT formalism allows 
Higgs boson couplings to be extracted via a much larger global fit.
This fit includes not only the basic observables above but also
additional observables of the reaction $\ee\to Zh$, as well as  observables of
electroweak precision physics and $\ee\to W^+W^-$.  These latter
measurements can be included because the EFT Lagrangian is the
complete Lagrangian and applies to all processes
 that occur in $\ee$ annihilation.
Though the number of free parameters is significantly increased,
it turns out that each parameter can be well 
controlled experimentally. Then the EFT  
improves significantly the measurement of Higgs boson couplings. 
The detailed strategy is explained in Section 3 and
 in \cite{Barklow:2017suo,Barklow:2017awn}. 
The results of Higgs boson coupling precisions 
based on the fitting program used in \cite{Barklow:2017suo,Barklow:2017awn} 
are given in Table \ref{tab:higgscouplings}.    These results are
illustrated and compared to the projections of Higgs coupling
uncertainties at HL-LHC from the ATLAS
 experiment~\cite{H2aaLHC} in Fig.~\ref{fig:Higgssummary}.

While the EFT coefficients
parametrize shifts in the Higgs couplings from massive new particles,
the fit that we use also allows Higgs decays to new particles lighter than
$m_h/2$, manifested both as invisible Higgs decays and as other modes
of exotic decay.\footnote{It is very conservative at an $\ee$ collider
  to allow that as many as 1\% of Higgs decays will remain
  unrecognized as distinct processes.  However, we do allow this
for the purpose of this fit.}
 The small difference with the numbers in \cite{Barklow:2017suo}
comes from the different luminosity sharing among $(-+,+-,--,++)$
assumed in the run plan presented in Section 2.

There are many interesting features
 in the Higgs boson coupling determination in the  EFT formalism.
It is worth emphasizing a few of them:
\begin{itemize}
\item
A unique role is played by the 
inclusive Zh cross section, $\sigma_{Zh}$, enabled by the recoil mass
technique.  This remains the key element in the determination of the
absolute normalization of all Higgs boson couplings.   This freedom is
mainly captured by the parameter $c_H$ of 
the EFT formalism. 
\item
The ratio of partial widths $\Gamma(h\to WW^*)/\Gamma(h\to ZZ^*)$ 
is determined very precisely, to $<0.15\%$, 
mainly thanks to the constraints imposed on the EFT Lagrangian by
$SU(2)\times U(1)$ gauge symmetry.  In Table~\ref{tab:higgscouplings},
we give values for effective couplings $g(hWW)$, $g(hZZ)$ defined to
be proportional to the square roots of the partial widths.
We see that  $g(hWW)$ can be determined as precisely as $g(hZZ)$
without relying on the  $\sigma_{\nu\nu h}$ 
measurement using the  $WW$ fusion process. 
This essentially solves the largest problem in measurement
 of Higgs boson couplings at $\sqrt{s}=250$ GeV.
Note that once $g(hWW)$ and $g(hZZ)$ are determined, 
$\Gamma_h$ and all other couplings $g(hA\bar A)$ are 
determined straightforwardly  using \leqn{eqn:TotalWidth} and
\leqn{eqn:PartialWidthA}.

\begin{figure}
\begin{center}
\includegraphics[width=0.90\hsize]{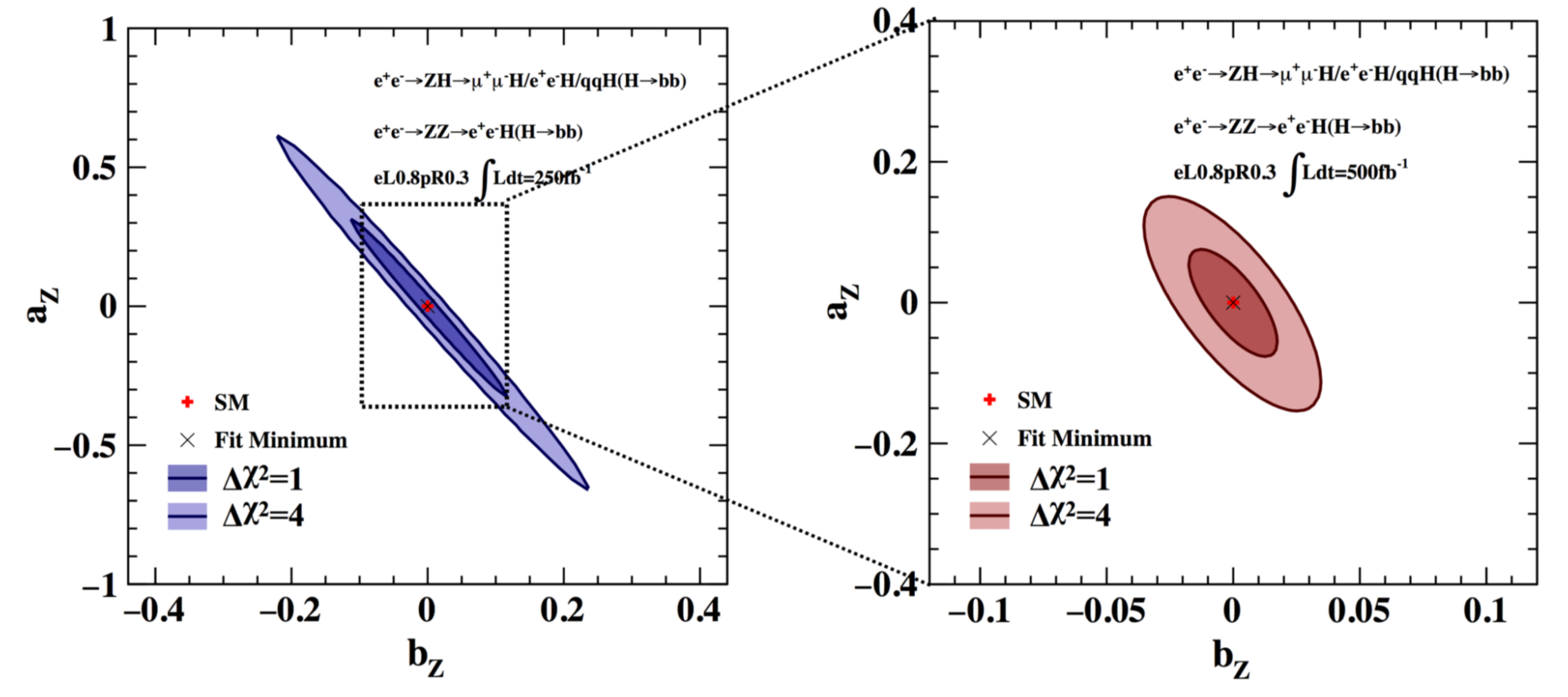}
\end{center}
\caption{Contour plots for $a_Z$ versus $b_Z$ from \cite{Ogawa}: for $\sqrt{s}=250$ GeV with 250 fb$^{-1}$ (left);
for $\sqrt{s}=500$ GeV with 500 fb$^{-1}$ (right).}
\label{fig:AandB}
\end{figure}
\item
In $e^+e^-\to Zh$, new observables making use of both cross section
and angular information are included.
The information in these observables is contained in two parameters
$a_Z$, $b_Z$
for each initial polarization  state \cite{Ogawa}.  The parameter $a_Z$ contains
the $\eta_Z$ term and is essentially identical for the polarizations
$\ELER$ and $\EREL$.  The parameter $b_Z$ contains $\zeta_Z$ or
$c_{WW}$ and also
an effect of photon-$Z$ mixing that is predicted by the EFT Lagrangian
to depend on $c_{WW}$ and related parameters. 
Estimates of the accuracy of the $a_Z$ and $b_Z$ measurements for
individual polarization states, based on
full detector simulation 
described in \cite{Ogawa}, are given 
in the Appendix of \cite{Barklow:2017suo}.
It is found that accuracies of the $a_Z$ and $b_Z$ determinations at 250 GeV 
are rather limited compared to that at 500 GeV; see Fig. \ref{fig:AandB}.
 Luckily, there are other powerful means
to help constrain $c_{WW}$.
\begin{figure}
\begin{center}
\includegraphics[width=0.60\hsize]{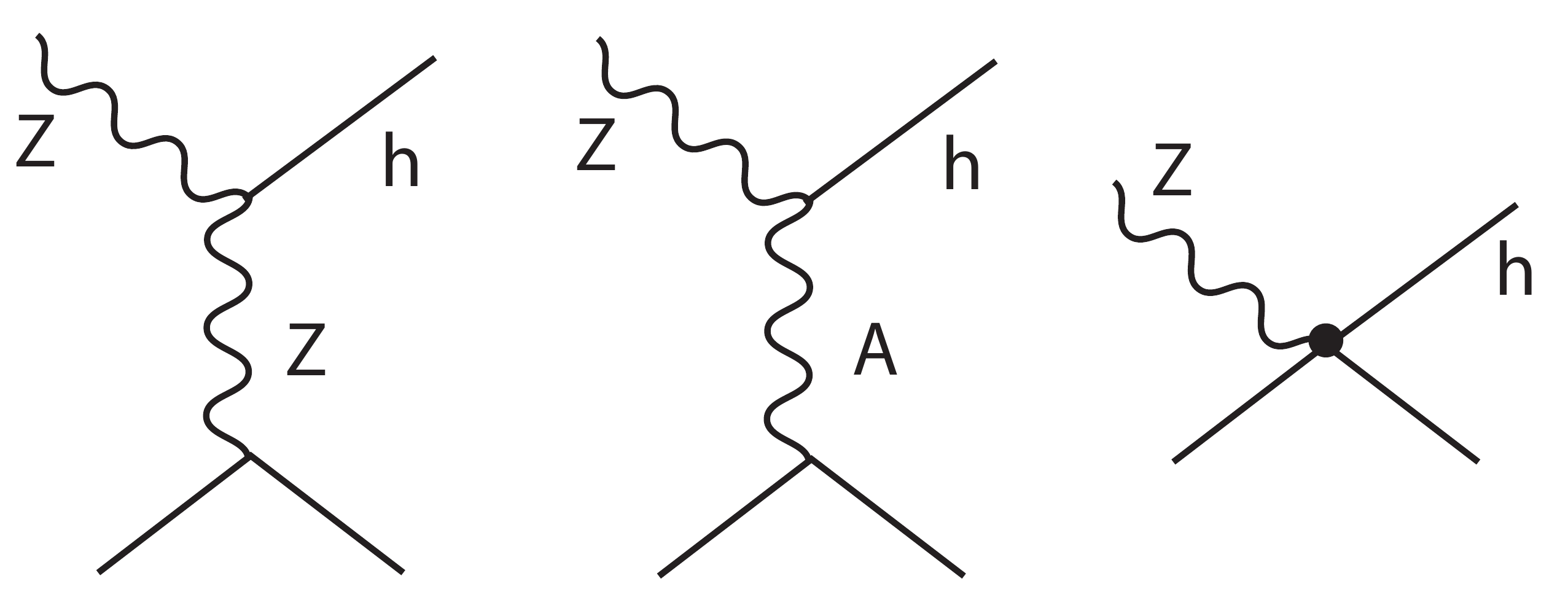}
\end{center}
\caption{Feynman diagrams contributing to the amplitudes for $\ee\to Zh$.}
\label{fig:Zhdiagrams}
\end{figure}
\item
The photon-Z mixing effect that contributes to $b_Z$ leads an additional
diagram  for $\ee\to Zh$ with  an $s$-channel photon instead of a $Z$.
 This diagram is shown in 
Fig.~\ref{fig:Zhdiagrams} along with a  third diagram that  arises from
dimension-6 operator vertices.    The interference between the
first two diagrams is constructive for $\ELER$ and destructive for
$\EREL$.   Since the mixing effect depends strongly on $c_{WW}$, this
EFT coefficient
can be constrained very well using measurement of the polarization 
asymmetry in $\sigma_{Zh}$.
Note that this polarization asymmetry in $\sigma_{Zh}$ can be
determined from   
using $\sigma_{Zh}\cdot BR$ measurements (which can be done with
hadronic decay modes of the $Z$) as well as from inclusive
cross section measurements (which are dominated by leptonic decays of
the $Z$).  This allows more of the total data set to be used to
constrain $c_{WW}$. The overall effect of beam polarizations on 
the determination of 
Higgs boson couplings can be found in Table 4 of \cite{Barklow:2017suo}.
\item   The decays
$h\to\gamma\gamma$ and $h\to Z\gamma$ are loop-induced in the SM,  but
receive corrections at the tree level from dimension-6 operators.  Thus,
$\Gamma_{\gamma\gamma}$ and $\Gamma_{Z\gamma}$ are very sensitive 
to the operator coefficients  $c_{WW}$, $c_{WB}$ and $c_{BB}$, the same set of operators that determine
$\zeta_A$, $\zeta_{AZ}$, $\zeta_Z$ and $\zeta_W$. The measurements
of $BR_{ZZ}/BR_{\gamma\gamma}$ and $BR_{\gamma Z}/BR_{\gamma\gamma}$ from the HL-LHC turn out
to be very helpful, 
providing tight constraints on two linear combinations of $c_{WW}$, $c_{WB}$ and $c_{BB}$,
even though the projected accuracy for the $Z\gamma$ decay is only $31\%$ \cite{H2aaLHC}. 
It will be  interesting to study 
whether any observable at ILC can measure the $hZ\gamma$ coupling
directly to still  better accuracy.
\item
The Triple Gauge Couplings (TGCs) measured in $e^+e^-\to W^+W^-$ play
a very important role in fixing three of the 17 relevant EFT
coefficients. So it is important that ILC will dramatically improve
the measurement of TGCs over what has been accomplished at LEP2 and
LHC.  We will discuss this measurement in Section 8 below. 
\item
The rightmost diagram in Figure \ref{fig:Zhdiagrams} is induced by
contact interactions from dimension-6 operator coefficients  that correct the
$Z$-lepton vertices measured from precision electroweak observables. 
These parameters appear in 
$\sigma_{Zh}$ with very large coefficients, of order $2s/m_Z^2\sim
15\, (60)$ at $\sqrt{s}=250\, (500)$ GeV.   It turns out that the
constraints on these coefficients is improved over that from the
current precision electroweak measurements by the comparison of Higgs
cross sections at 250 and 500 GeV.\footnote{The fit described here
  uses only the current uncertainties in precision electroweak
  measurements, except for an improvement in the uncertainty
 in $\Gamma_W$ to 0.1\% expected from ILC measurements of final states
 in $\ee\to
  W^+W^-$.} Alternatively, the EFT fit would
be assisted by improvement of precision electroweak measurements,
either by direct $\ee$  running at the $Z$ pole or by measurements of
the polarization asymmetry of 
the radiative return process $\ee\to Z \gamma$.  This is another topic
that needs further investigation. 
\end{itemize}

\subsection{Measurement of the Higgs boson mass and CP}

The uncertainty in the  Higgs boson mass ($\delta m_h$) is a 
source of systematic error for predictions of Higgs boson couplings. 
In most cases, 
 $\delta m_h\sim 0.2\%$ would be already sufficient, but this is not
 true
 for $h\to ZZ^*$ or
$h\to WW^*$. It has been pointed out in  \cite{Almeida:2013jfa} that 
\beq
\delta_W =6.9\cdot\delta m_h,~~~~\delta_Z =7.7\cdot \delta m_h,
\eeq{eqn:MassH}
where $\delta_W$ and $\delta_Z$ are the 
relative errors for $g(hWW)$ and $g(hZZ)$ respectively. 
At the 250~GeV ILC, the Higgs boson mass can be measured very precisely, 
with an accuracy of 14 MeV using the leptonic recoil channel 
as shown in Fig. \ref{fig:Sigmah} (left).  This results in 
systematic errors for $\delta_W$ and
$\delta_Z$ of 0.1\%. The study of the new beam parameters discussed in Section~2,
which would increase the beamstrahlung effect significantly, should 
pay attention to this issue.

At the 250~GeV ILC,  Higgs CP properties can be  probed via the
$h\tau\tau$ coupling, 
\beq
\Delta{\cal L}_{h\tau\tau}=-\frac{\kappa_\tau y_\tau}{\sqrt{2}}h{\tau^+}(\cos\phi+i\sin\phi\gamma_5)\tau^-
\eeq{eqn:CPHtautau}
and the $hVV$ coupling
\beq
\Delta{\cal L}_{hZZ}=\frac{1}{2}\frac{\tilde{b}}{v}hZ_{\mu\nu}\tilde{Z}^{\mu\nu}.
\eeq{eqn:CPHZZ}
The CP phase $\phi$ in \leqn{eqn:CPHtautau} can be measured with an accuracy of $3.8^\circ$
\cite{Jeans:2016}, and $\tilde{b}$  in \leqn{eqn:CPHZZ} can be
measured with an accuracy of 
0.5\% \cite{Ogawa}.
The observation of even a  small admixture of CP-odd coupling for the
Higgs boson would indicate physics beyond the Standard Model, and
might give a clue to the CP violation required in models of
electroweak baryogenesis.

\section{Comparison of the ILC capabilities for the Higgs boson to the predictions of new
  physics models}

Now that we have presented the expectations for the accuracy of ILC measurements of the Higgs boson couplings, it is important to ask whether these expectations are strong enough to search for new physics beyond the reach of direct searches at the LHC.   We will discuss that point in this section.
First, we will survey models of new physics that affect the Higgs boson, review
the diversity of models under consideration, and present the effects on the Higgs couplings predicted in the various types of models. Then we will present a 
representative sample of specific model points that demonstrate the power of the ILC measurements.

\subsection{Models of electroweak symmetry breaking and the Higgs field}

Our present understanding of the breaking of the $SU(2)\times U(1)$ gauge symmetry of the SM is crude and unsatisfactory.   This point is, suprisingly, more easily grasped by condensed matter physicists than by particle physicists.  Condensed matter physicists are familiar with the history of superconductivity, for which the basic understanding developed in two stages.  In 1950, Landau and Ginzburg wrote a phenomenological theory of superconductivity that was, in fact, the model for the theory of the Higgs
 field~\cite{Landau}. This model was successful and even predictive of many aspects of superconductivity, but, in this model,  the basic fact of the phase transition to superconductivity was put in by hand.  Only later, in 1957, did Bardeen, Cooper, and Schrieffer (BCS)  create a fundamental theory of superconductivity based on pairing of electrons in a metal~\cite{BCS}. The BCS 
theory was not only a conceptual improvement but also predicted many new features of superconductivity that were beyond the reach of the
 Landau-Ginzburg description.  In particle physics, we are now at the Landau-Ginzburg stage~\cite{Wells:2016exe}.   The difference from the condensed matter situation is that the interactions that drive electroweak symmetry breaking and lead to the phase transition must be new interactions, outside the SM, that have not yet been discovered.  Thus, the exploration of the Higgs field offers the opportunity to discover genuinely new interactions of nature.

Many features of the SM argue that it cannot be a fundamental solution for 
electroweak symmetry breaking.   For example, there are good reasons to believe that a scalar boson cannot be light (compared to Planck scale, for example) and give mass to all other particles in the SM without aid from other---as yet undiscovered---particles and interactions. These additional particles necessarily interact with the Higgs boson and can change the expectations for Higgs couplings to SM states. 

Theories of physics beyond the SM are constructed to solve one or more outstanding problems that the SM does not address. 
They might attempt to explain the
 low mass of the Higgs boson without large fine-tunings as discussed above; 
they might posit dark matter candidates; they might  explain the baryon asymmetry of the universe; they might unify the SM forces. In this space of theories there are many that can produce experimentally accessible
 non-SM signals to be discovered in the near term --- some through direct searches of particles at the LHC's high-energy frontier, and others through a 
myriad of other experiments currently running or planned for the near future. 
Among these probes, though,   precision measurement of the Higgs couplings is one of the most powerful. The reasons for this are two-fold. First, the Higgs sector is where we expect the most new interactions in many beyond 
the SM theories.  Second, measurements in the Higgs sector have great room for improvement over current precision levels that can reveal new physics effects for beyond the SM theories that no other experiments could access. 

\begin{table}[t]
\begin{center}
\begin{tabular}{lccc}
     & $\Delta g(hVV)$ & $\Delta g(ht\bar t)$ & $\Delta g(hb\bar b)$ \\
\hline
Composite Higgs & 10\% & tens of \% & tens of \% \\
Minimal Supersymmetry & $<1\%$ & 3\% & tens of \% \\ 
Mixed-in Singlet & 6\% & 6\% & 6\% \\
\end{tabular}
\caption{ Estimated maximum deviations of Higgs couplings to various SM states allowed by three different scenarios of physics beyond the SM. The assumption is that no new physics associated with electroweak symmetry breaking is found at the HL-LHC  ($3\, {\rm ab}^{-1}$ at $\sqrt{s}=14\, {\rm TeV}$), and thus Higgs coupling measurements are the only potential signal for new physics. Adapted from~\cite{Gupta:2012mi}.}
\label{tab:grw}
\end{center}
\end{table}

Models that address the problems just listed can be constructed using many different approaches.   A survey of these approaches, and estimates of the maximum possible effect on the Higgs boson couplings, was given
 in~\cite{Gupta:2012mi}.   Table~\ref{tab:grw} lists three important classes of models  and the corresponding estimates for maximal deviations in different couplings of the Higgs boson. 

One class of models builds on the analogy with superconductivity.  The order parameter for the Landau-Ginzburg potential of superconductivity turned out to be a composite state of electrons (Cooper pairs). It has been suggested that the Higgs boson state is also composite.  If this is true, it has the potential to explain the large hierarchy between the Higgs mass and the Planck scale.  A collection of some of the simplest approaches along this line leads to potentially large deviations of Higgs boson couplings to SM states compared to the expected measurement accuracies from the ILC.

\begin{figure}
\begin{center}
\includegraphics[width=0.90\hsize]{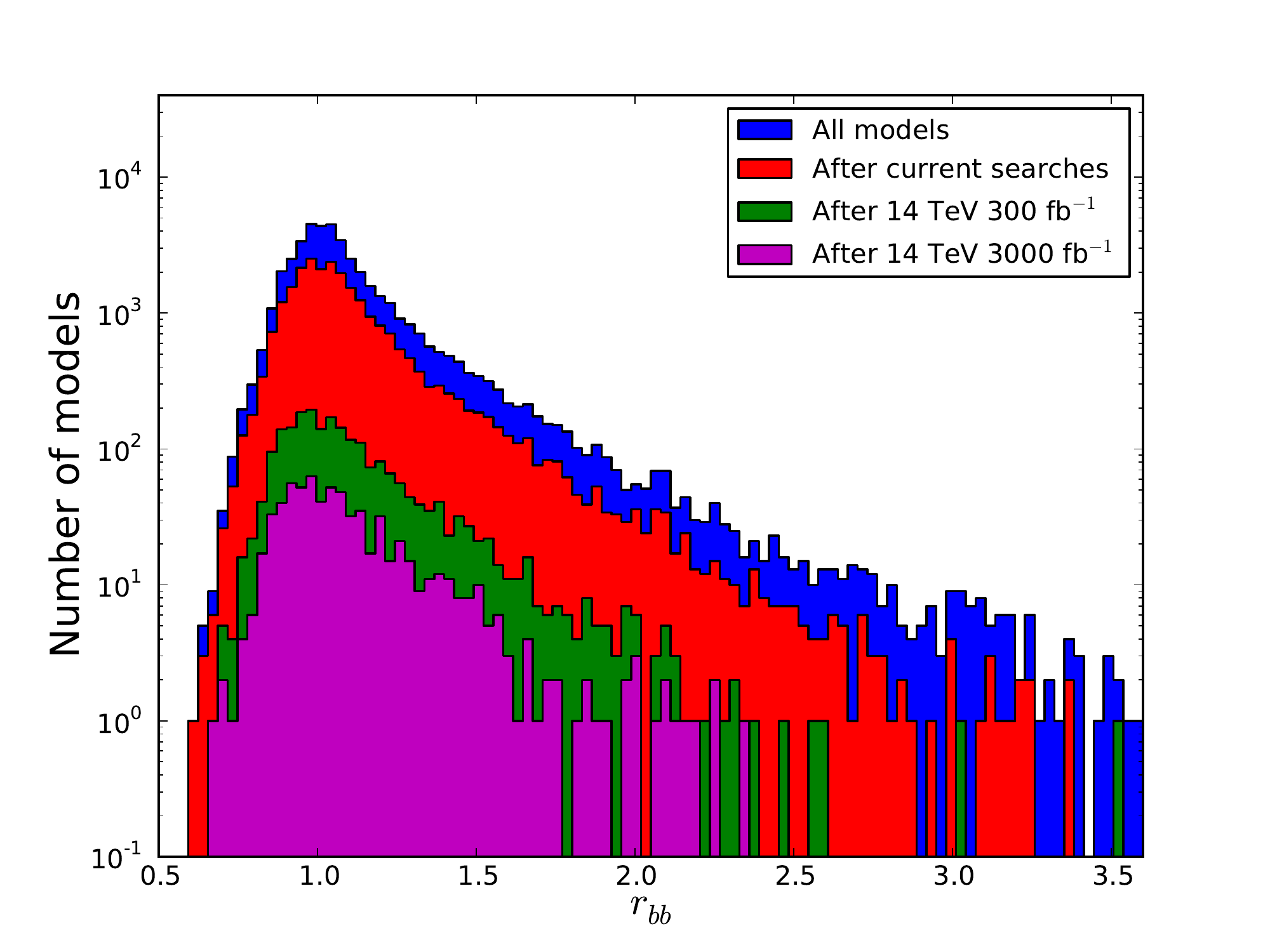}
\end{center}
\caption{Histograms of the ratio $r_{bb}=\Gamma(h\to \bar bb)/\Gamma(h\to \bar bb)_{\rm SM}$ within a scan of the approximately 250,000 supersymmetry parameter sets after various stages of the LHC, assuming the LHC does not find direct evidence for supersymmetry.  The purple histogram shows parameter points  that would not be discovered at future upgrades of the 
LHC (14 TeV and $3\, {\rm ab}^{-1}$ integrated luminosity).  From~\cite{Cahill-Rowley:2013vfa}. }
\label{fig:PMSSM}
\end{figure}

A different class of models makes use of supersymmetry.
Supersymmetry posits a symmetry between bosons and fermions that not only could explain the Higgs boson mass with respect to the Planck mass, but it could also be the source of dark matter, and it could be the key ingredient that enables the unification of forces at the high scale~\cite{Martin:1997ns}. The symmetry requirements of supersymmetry require the introduction of two Higgs bosons -- one that gives mass to up-type fermions and one that gives mass to down-type fermions. The two Higgs doublets mix and leave one CP-even eigenstate light, which is identified with the 125 GeV Higgs boson ($h$). It is straightforward to derive that this light boson $h$ has couplings identical to those of the SM Higgs boson except for small deviations that are induced by mixings with the extra Higgs states and loop corrections involving the 
superpartners and the heavy Higgs bosons. These deviations of couplings can be well above $10\%$ in the case of Higgs coupling to $b$ quarks, even if no superpartner is ever found at the LHC in all its planned upgrade phases~\cite{Gupta:2012mi}. This is illustrated nicely by Fig.~\ref{fig:PMSSM}, where the authors scanned over hundreds of 
thousands of MSSM supersymmetric points~\cite{Cahill-Rowley:2013vfa}. They showed that many sets of parameters in the MSSM can never be found at the LHC but would be easily discernible through precision measurements at the ILC. 

A third class of models postulates  additional scalar fields.  After all, there are many fermions, and there are many vector bosons.
 Multiple scalars are already required within supersymmetry, where in addition to scalar superpartners we stated that two Higgs bosons are required. 
But there are many more ideas of beyond the SM physics that incorporate several scalar bosons but do not cause ill effects elsewhere, by, for example, inducing too large flavor changing neutral currents.
 These multi-Higgs doublet models are classified as  type I (in which 
one Higgs gives mass to fermions, and the other does not), type II (in which 
one Higgs gives mass to  up fermions only and one to down fermions only), and type X and Y models
 (with more complicated discrete symmetries that protect flavor observables)~\cite{Kanemura:2014dea}. 

\subsection{Comparisons of models to the ILC potential}

\begin{table}
\begin{center}
\begin{tabular} {llccccccccc}
 &Model  & $b\bar b$ &       $c\bar c$    & $gg$  &  $WW$ & $\tau\tau$ &   $ ZZ$  &
                      $\gamma\gamma$    & $\mu\mu$   \\ \hline
1&MSSM~\cite{Cahill-Rowley:2013vfa}  &  +4.8 &-0.8  &  - 0.8& -0.2 & +0.4 &  -0.5 &
                    +0.1   &   +0.3\\
2&Type II 2HD ~\cite{Kanemura:2014dea} &+10.1  & -0.2 &-0.2 & 0.0&+9.8 &0.0
                                                             &+0.1  &+9.8\\
3&Type X 2HD ~\cite{Kanemura:2014dea} &-0.2  &-0.2 &-0.2 &0.0&+7.8
                                                   &0.0 &0.0 &+7.8\\
4&Type Y 2HD ~\cite{Kanemura:2014dea} &+10.1  &-0.2 & -0.2  & 0.0 &-0.2 &
                                                 0.0 &0.1 &-0.2\\
5&  Composite Higgs~\cite{Contino:2006qr} &-6.4 & -6.4 & -6.4& -2.1& -6.4&
                                                   -2.1& -2.1& -6.4\\
6&Little Higgs w. T-parity~\cite{Hubisz:2005tx} &
                   0.0 & 0.0  &   -6.1   &   -2.5  &  0.0  &  -2.5
                                             &  -1.5 &    0.0 \\
7&Little Higgs w. T-parity~\cite{Chen:2006cs} &
                    -7.8 & -4.6  &   -3.5   &   -1.5  &  -7.8   &   -1.5
                                             &  -1.0  &-7.8 \\
8&Higgs-Radion~\cite{Hewett:2002nk} &  -1.5  &  - 1.5   & +10. & -1.5 & -1.5 & 
                  -1.5 &    -1.0 & -1.5 \\
9&Higgs Singlet~\cite{DiVita:2017eyz} &  -3.5 &  -3.5 & -3.5& -3.5 &
                                                                     -3.5 & -3.5 & -3.5& -3.5 \\
\end{tabular}
\caption{Percent deviations from SM for Higgs boson couplings to SM states in various new physics models. These  model points are unlikely to be discoverable at 14 TeV LHC through new particle searches even after the high luminosity era ($3\, {\rm ab}^{-1}$ of integrated luminosity). From~\cite{Barklow:2017suo}.}
\label{table:models}
\end{center}
\end{table}

\begin{figure}
\begin{center}
\includegraphics[width=0.85\hsize]{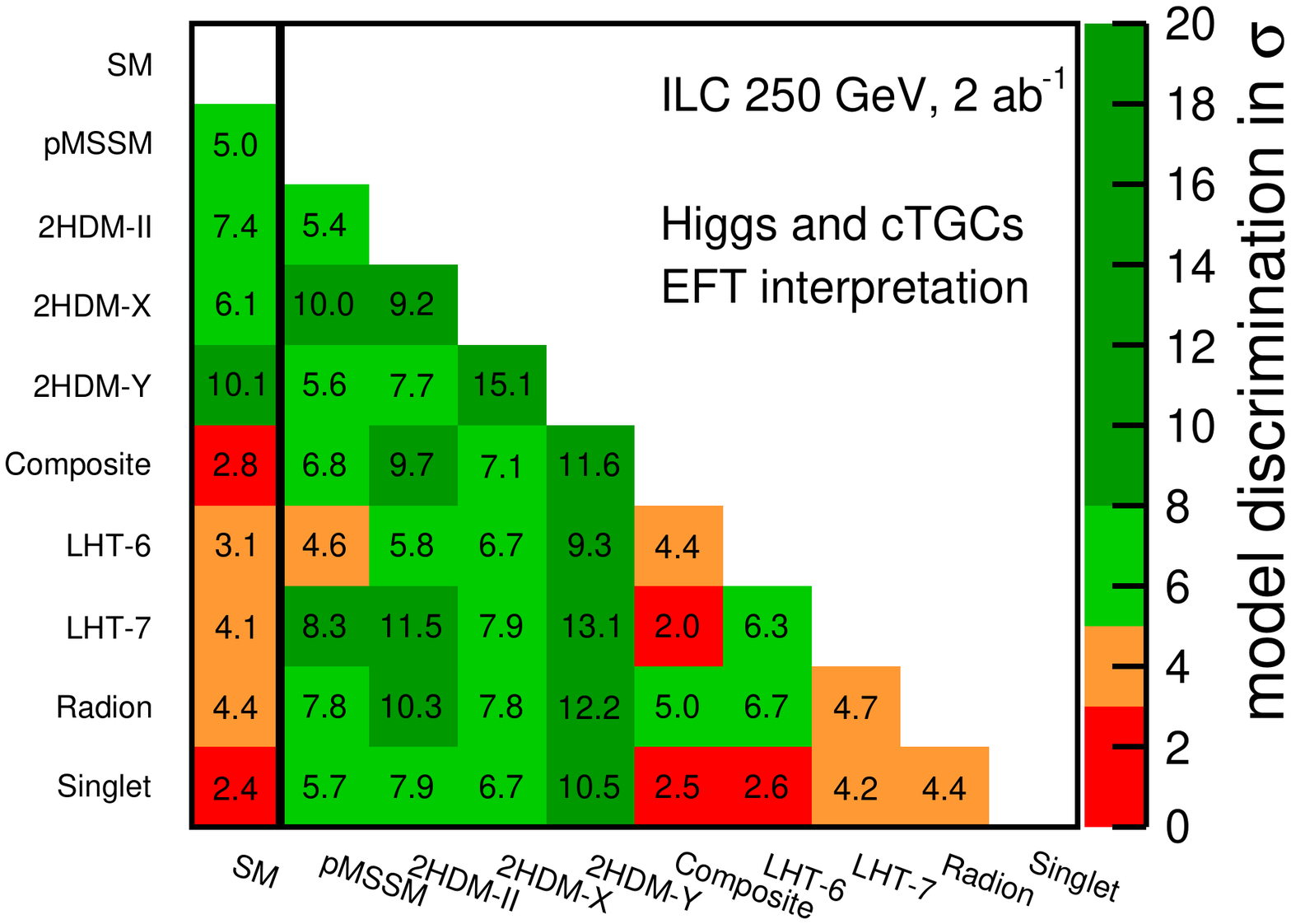} \\
\includegraphics[width=0.85\hsize]{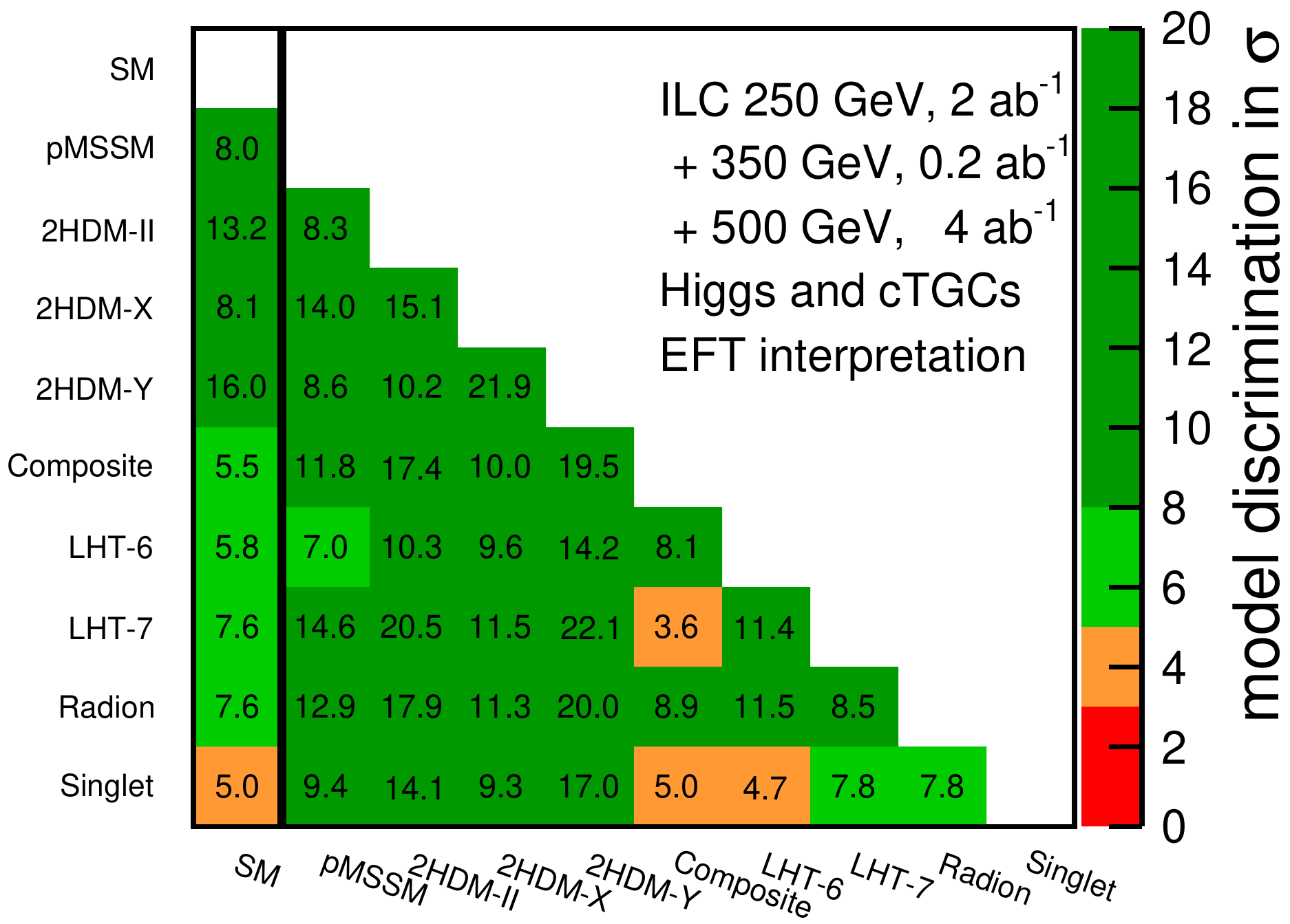}
\end{center}
\caption{Graphical representation of the $\chi^2$ separation of  the
 Standard Model and the models 1--9 described in the text:  (a) with
 2~ab$^{-1}$  of data at the ILC at 250~GeV; (b) with
 2~ab$^{-1}$  of data at the ILC at 250~GeV plus 
4~ab$^{-1}$  of data at the ILC at 500~GeV. Comparisons in orange have
above 3~$\sigma$ separation; comparison in green have above 5~$\sigma$
separation; comparisons in dark green have above 8~$\sigma$
separation. From \cite{Barklow:2017suo}, with slight modifications to account
for the beam polarization scheme in Section~2. }
\label{fig:chisq}
\end{figure}

All of these  ideas lead to models with deviations from the SM expectations
of the couplings of the 125 GeV Higgs boson to SM states. Table~\ref{table:models} collects a set of models of new physics based on the ideas described in the previous section and on several additional  ideas of interest to theorists.  For each model, we chose a representative parameter point for which the predicted new particles would be beyond the  reach of the 14 TeV LHC  with the full projected data set.  The deviations
 of Higgs couplings from the SM
expectations at these representative model points are listed in the Table.
  (For details, see \cite{Barklow:2017suo} as well as the papers cited in Table~\ref{table:models}.)   These examples  illustrate diverse possibilities for models with significant deviations of the Higgs couplings from the SM expectation that would be allowed even if the LHC and other experiments are not able to discover the corresponding new physics beyond the SM. We should make  clear that the quantitative statements to follow  refer to these 
particular models at the specific parameter points shown in the Table.   Figure~\ref{fig:chisq} shows graphically the ability of ILC measurements to distinguish the Higgs boson couplings in the models in the Table  from the SM expectations and from the expectations of other models.   Each square shows relative goodness of fit for the  two models in units of $\sigma$.  The top figure is based on the covariance matrix
 from the 250~GeV stage of the ILC, corresponding to the second column of Table~\ref{tab:higgscouplings}.  The bottom figure reflects
 the full ILC program with 500 GeV running, corresponding
 to the fourth column of Table~\ref{tab:higgscouplings}.  
 It is noteworthy that,
once it is 
known that the Higgs boson couplings deviate
 significantly from the SM predictions, the pattern of the deviations is
 characteristic for each model and 
distinguishable from the patterns predicted by other models
 in the set.

The evidence for significant deviations in the Higgs boson couplings would demonstrate that there is new physics beyond the SM that affects the Higgs field.  The observation of the  pattern of deviations would give us information on the properties of this new physics and point the way to further model-building and experimental exploration.   This is a route to a deeper understanding of nature that the ILC offers us.




\section{Invisible and exotic Higgs decays}

In addition to the expected decays of the Higgs boson whose analysis was 
discussed in the previous two sections, the Higgs boson could also have additional decay modes that are not predicted by the SM.    The ILC at 250~GeV will accumulate a data set containing half a million Higgs bosons tagged by recoiling $Z$ bosons.  This will provide an ideal environment to search for any possible final state of Higgs decay.  

Exotic decays of the Higgs boson are expected in many theoretical models.  An attractive way to model the dark matter of the universe is to assume the 
existence of a 
 ``hidden sector'' 
consisting of one or more fields with no SM gauge charges. Since particles of  a hidden sector do  not couple through  gauge forces, their interactions with SM particles are highly model-dependent and can be very feeble.  Such particles can be consistent with all existing experimental constraints even if their masses are well below the weak scale.   If some of the hidden-sector particles are stable, these could make up  the observed dark matter.  For example, in the ``Strongly-Interacting Massive Particle" (SIMP) scenario~\cite{Hochberg:2014dra,Hochberg:2014kqa}, dark matter consists of mesons produced by confinement of a QCD-like gauge group in the dark sector. A light hidden sector also appears in well-motivated theoretical models of electroweak symmetry breaking   such as the  ``Twin Higgs" model~\cite{Chacko:2005pe}.  Hidden-sector particles have also been invoked as an explanation of the apparent discrepancy between
 the experimental and theoretical values of the anomalous magnetic moment of the muon, and a number of other experimental anomalies. In light of this, there is strong interest in experimental searches for these particles, and a number of approaches are currently being pursued or studied~\cite{Alexander:2016aln,Battaglieri:2017aum}. 

To connect the hidden-sector particles to initial states with Standard Model particles, it is necessary to add a term to the Lagrangian that connects these sectors.   There are precisely three dimension-4 operators that can make this connection:
\beq
     \eps\, B_{\mu\nu}{\hat F}^{\mu\nu} \ , \qquad  \eps\, |\varphi|^2 |\hat S|^2\ , 
\qquad \eps \, L^\dagger \cdot \varphi \hat N \ ,
\eeq{portals}
where $B_{\mu\nu}$ is the $U(1)$ field strength, $\varphi$ is the Higgs doublet, and 
$L$ is the lepton doublet of the SM and fields with hats are in the hidden sector.   These are called the ``gauge portal'', ``Higgs portal'', and ``neutrino portal'', respectively.  Note that the neutrino portal also involves the Higgs field. Almost all of the attention in the 
reports \cite{Alexander:2016aln,Battaglieri:2017aum} is given to the gauge portal,
which can be studied with low-energy fixed-target experiments, among other techniques.  This leaves open a wealth of other possibilities, especially if the hidden sector particles have masses above a few GeV.

Decays of the Higgs boson offer a unique opportunity for very sensitive searches for a light hidden sector using the Higgs and neutrino portals.  The SM Higgs width is tiny, $\Gamma_h/m_h \simeq 3\cdot 10^{-5}$. Thus, the branching fraction of Higgs decays to hidden-sector states could be sizable even if its couplings to such states are rather small. 
	
Signatures of Higgs decays into the hidden sector are model-dependent. One possibility is that the hidden-sector particles are stable or sufficiently long-lived that they do not decay inside the detector. Since interactions between hidden-sector particles and ordinary matter are extremely weak, they will escape the detector unseen, resulting in an ``invisible Higgs decay" signature. Experiments at an electron-positron collider have excellent sensitivity to this signature, due to their ability to tag Higgs bosons using the 
recoil mass technique.  The 250~GeV ILC is expected 
to be sensitive to invisible Higgs decays with branching ratios as small as 
$0.3$\%~\cite{Barklow:2017suo}, a factor of 20 
below the expected HL-LHC sensitivity.

Another interesting possibility is that the hidden-sector particles decay inside the detector. If the decay occurs purely within the hidden sector, such final states would remain invisible. On the other hand, if the decay products include SM particles\footnote{Even if all couplings between the hidden sector and the SM are small, such decays may occur with significant probability, {\it e.g.} in cases when no competing decays within the hidden sector are kinematically available.}, they are potentially observable as ``exotic" Higgs decay modes. A large variety of decay topologies and specific final states are possible; a systematic discussion can be found in the recent overviews of Higgs exotic decays~\cite{deFlorian:2016spz,Curtin:2013fra}. Two simple and theoretically well-motivated examples are:

\begin{enumerate}
	
	\item $f\bar{f}+\met$, where $f$ is an SM fermion. For example, in SUSY models with an extra gauge-singlet scalar $s$, such as the NMSSM, this final state arises from the decay chain $h\to\neu{1}\neu{2}$, $\neu{2}\to s\neu{1}$, $s\to f\bar{f}$, with either on-shell or off-shell $s$. The flavor of $f$ is dictated by the couplings of $s$ to quarks and leptons, which are highly model-dependent.   If the connection to the hidden sector is through the neutrino portal, the neutrino will provide missing energy even if the mediator fermion labelled $\hat N$ in \leqn{portals} produces only visible final particles.

	\item $(f\bar{f})(f^\prime\bar{f}^\prime)$, where $f$ and $f^\prime$ are SM fermions, and brackets indicate a resonant pair. These final states arise from a decay chain $h\to aa$, $a\to f\bar{f}$, where $a$ is a gauge-singlet scalar particle, for example a composite of a confining hidden-sector gauge group in Twin Higgs models. Again, the fermion flavors involved in these decays are highly model-dependent. 
	
\end{enumerate}

Experiments at the (HL-)LHC will have excellent sensitivity to exotic Higgs decay modes to  electrons, muons, or photons. However, final states involving quarks or tau leptons are very challenging at the LHC. The ILC at 250~GeV offers a perfect environment to search for such final states, due to low QCD backgrounds and Higgs tagging with recoil-mass technique. The paper \cite{Liu:2016zki} estimated the sensitivity of the 250~GeV ILC with 2 ab$^{-1}$ integrated luminosity to the exotic decay topologies listed above. In the $f\bar{f}+\met$ channels, with $f=j, b$ or $\tau$, the ILC will be sensitive to branching ratios in the $10^{-4}-10^{-3}$ range, {\it vs.} a projected sensitivity of at best 20\% at the LHC. For the $(f\bar{f})(f^\prime\bar{f}^\prime)$ topology, the improvement is equally dramatic: for example, branching ratios of the $(b\bar{b})(b\bar{b})$, $(c\bar{c})(c\bar{c})$, $(j\bar{j})(j\bar{j})$ channels will be probed down to the level of $10^{-3}$, improving the LHC sensitivity by two orders of magnitude.

\section{Opportunities for discovering direct production of new particles}

Although the LHC experiments have carried out extensive searches for new particles, these searches have well-recognized limitations.   The LHC exclusions are strongest for particles produced by QCD interactions and are less powerful for particles produced through electroweak processes, which have  smaller cross sections.  Discovery at the LHC is especially difficult if the new particles decay with very small visible energy, for example, if a charged particle decays to a stable  neutral partner separated in mass by less than 15~GeV.   For particles of this type, the best current limits can still come from the LEP 2 experiments.

It is not clear {\it a priori} that ILC at 250~GeV offers a significant discovery reach beyond LEP 2.  The  center-of-mass energy of 250~GeV is only about 40~GeV above the highest energies
reached at LEP~2. This argument, however, overlooks three features of the ILC program. First,
the ILC run at 250~GeV offers  about 1000 times more integrated luminosity than collected at the highest energies by all 4 LEP experiments together ($\sim 250$\,pb$^{-1}$ per experiment in the year 2000 vs.\ $2$\,ab$^{-1}$).  Second, the ILC offers  polarized beams which, 
especially in the $(+-)$ configuration,  can suppress SM backgrounds by 1-2 orders of magnitude,
thereby increasing the sensitivity to rare BSM events.  Finally, the ILC detectors will profit  from 30 years of advances in technology, giving more than an order of magnitude better momentum and impact parameter resolutions, a factor 2 improvement in the jet energy scale, and considerably tightened hermeticity.

\begin{figure}
\begin{center}
   \hspace{-0.1\hsize} 
   \begin{subfigure}{.40\hsize}
      \includegraphics[width=\textwidth]{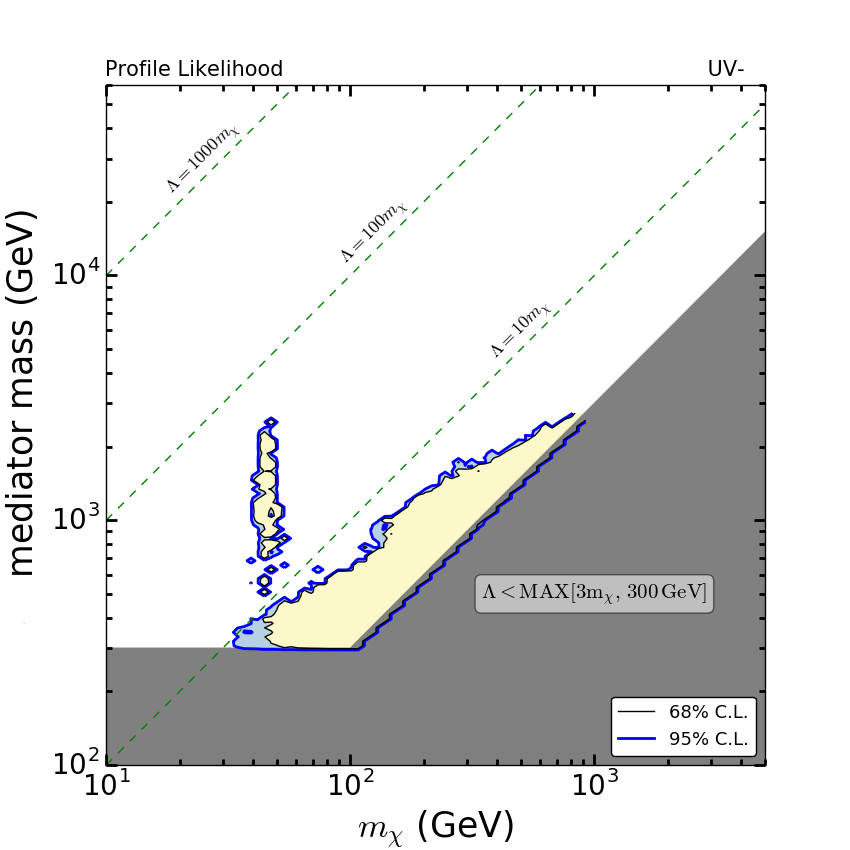}
         \subcaption{} \label{fig:WIMPlikeli}
   \end{subfigure} 
   \hspace{-0.001\hsize} 
   \begin{subfigure}{.58\hsize}
      \includegraphics[width=\textwidth, angle=-90]{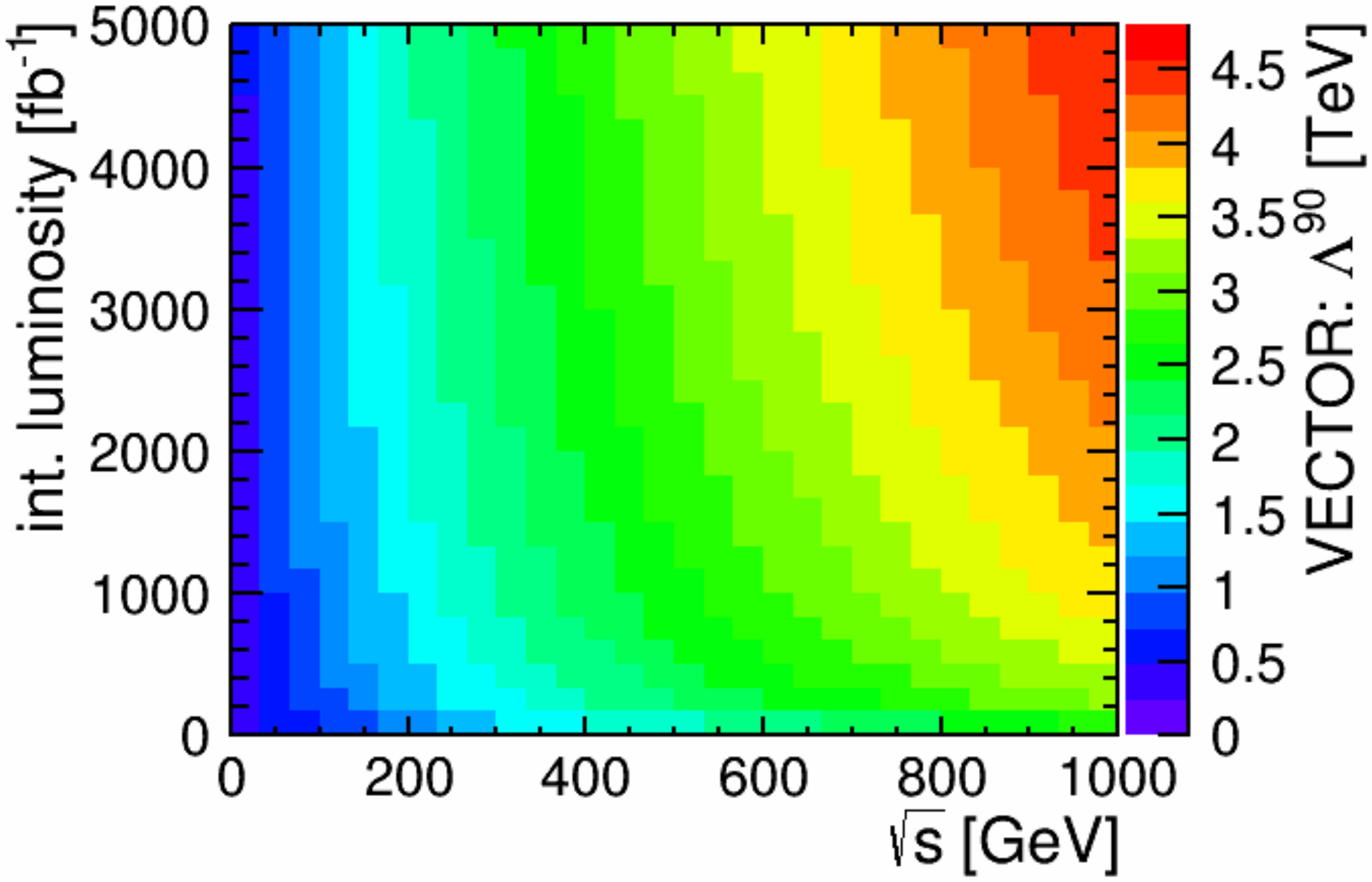}
         \subcaption{} \label{fig:WIMPreach}
   \end{subfigure}  
\end{center}
 \caption{Sensitivity of WIMP searches in the mono-photon channel, from  \cite{Habermehl:2017dxh}: (a) The yellow area indicates regions in WIMP parameter space which are not probed by current or future direct detection experiments or by searches at the (HL-)LHC. (b) New physics scale $\Lambda$ probed by mono-photon searches at the ILC as a function of center-of-mass energy and integrated luminosity.}
 \label{fig:WIMPs}
\end{figure}

Therefore, any search channel which was not kinematically but instead cross-section limited at LEP~2
offers significant discovery potential at the ILC, even at $250$\,GeV. One prominent example is  the search for additional light Higgs bosons.
Because the 125~GeV Higgs boson has couplings to $W$ and $Z$ close to those of the SM, additional bosons must have suppressed couplings to the $Z$ boson. These can be searched for as at LEP in specific decays modes, but probing couplings to the $Z$ boson at least one order of magnitude smaller. In addition, the much higher luminosity at the ILC will allow searches  for such particles independently of their decay mode via the recoil technique~\cite{Yan:2016trc}.

Even some SUSY searches were not yet kinematically limited at LEP.  For example, 
the LEP  lower limit on the mass of the supersymmetric partner of the $\tau$-lepton is only $26.3$\,GeV~\cite{Abdallah:2003xe} in the general MSSM, \ie, when allowing any mixing and any mass difference to the lightest SUSY particle.

Another interesting goal is the search for heavy sterile neutrinos. 
Improving the limits from LEP~1 on the mixing with the SM neutrinos at masses below $45$\,GeV would require an extended run at the $Z$ pole.  But such sterile neutrinos could also be produced directly together with a SM neutrino.  This process would show up as an apparent  deviation in the $W^+W^-$ production cross-section~\cite{Antusch:2016ejd,Liao:2017jiz}. In this case, 
the sensitivity is expected to expand the regime probed by LEP~2
 by at least an order of magnitude.

An important focus of new particle searches both at LHC and ILC is the search for WIMP pair production, which is observed at ILC in the mono-photon channel.
This search was studied in full simulation at 500~GeV, and the results of this study have been extrapolated to lower  center-of-mass energies~\cite{Habermehl:2017dxh}. The case of a singlet-like fermion WIMP  is illustrated in Fig.~\ref{fig:WIMPlikeli}.  Substantial regions of parameter space at masses below $\sim 120$\,GeV will remain even after 
a combined likelihood analysis including current and future direct detection as well as (HL-)LHC prospects. Figure~\ref{fig:WIMPreach} shows the new physics scale $\Lambda$ which can be probed by the ILC for the case of a vector-like fermion WIMP and a vector-like operator dominating its interactions with SM particles, as a function of the center-of-mass energy and the integrated luminosity, assuming a sharing between different beam 
helicity configurations of (40\%, 40\%, 10\%, 10\%) as 
described in Section 2. For $2$\,ab$^{-1}$ at $250$\,GeV, 
new physics scales up to $1.9$\,TeV can be probed.

\section{$\ee\to W^+W^-$ at 250~GeV}

Measurements of the $\gamma W^+W^-$  and $Z  W^+W^-$ triple 
gauge boson couplings (TGC's) test the $SU(2)\times U(1)$ gauge boson
self-coupling structure of the SM and probe BSM physics.  As for the
particle searches described in the previous section, the ILC at
250~GeV with 2~ab$^{-1}$ offers substantial improvement beyond the
results of LEP~2, which have not yet been surpassed by LHC.

The most general Lorentz invariant $\gamma W^+W^-$  or $Z  W^+W^-$
vertex contains 7 complex parameters, 
denoted by $g_1^V, g_4^V, g_5^V, \kappa_V, 
\lambda_V, \tilde{\kappa}_V, \tilde{\lambda}_V$,
$V=\gamma,Z$\cite{Hagiwara:1986vm}.  In total there are 14 complex
parameters to consider.  
At tree-level, in the SM,
$g_1^V=\kappa_V=1$ and all other parameters are zero.  SM 
radiative corrections are on the order 
of $2\times 10^{-2} M_Z^2/s$\ \cite{Arhrib:1996rj}.

 The primary focus of TGC studies is the search for 
modifications to the TGC's from BSM physics at 
energy scales well beyond the $\ee$ center-of-mass energy. 
As described in Section 3,   such physics is
parameterized by an effective Lagrangian with dimension-6 operators
that respects 
 $SU(2)\times U(1)$ gauge symmetry.   CP-conserving and CP-violating
 effects are separately measurable, with comparable accuracy.
   Here we will concentrate on the 
CP-conserving operators.  In this context,
only six real TGC parameters are relevant: $g_1^V,
\kappa_V, \lambda_V$ for $V = \gamma, Z$.  
 Furthermore, three $SU(2)\times U(1)$  constraints
\beqa
    g_1^\gamma & = &  1 \CR
    \kappa_Z & = & -(\kappa_\gamma -1)\tan^2\theta_W + g_1^Z \CR
    \lambda_Z & = & \lambda_\gamma 
\eeqa{tgcsu2xu1}
reduce the number of free
 parameters to three:  $g_1^Z, \kappa_\gamma, \lambda_\gamma$.

TGC's are measured at the ILC through the processes
$\ee\rightarrow W^+W^-$, and 
$e^-\gamma \rightarrow \nu_eW^-$, where the initial state $\gamma$ refers
to either a virtual or beamstrahlung photon.  Initial state beam
polarization can 
be used to disentangle
$\gamma W^+W^-$  couplings  from $Z  W^+W^-$.
The $W^-$ production polar angle $\Theta$ and the rest
frame fermion 
polar and azimuthal
angles, $(\theta^*,\phi^*)$ and $(\bar{\theta}\ ^*,\bar{\phi}\ ^*)$,
associated 
with the decays of the $W^-$ and $W^+$, respectively,
can be precisely measured.
The correlated distributions
of these five angles will be used as 
a polarization analyzer to separate out the multiple combinations of
transversely and 
longitudinally polarized $W^-$ and $W^+$ bosons.

In order to properly estimate the TGC sensitivity of the 
ILC at $\sqrt{s}=250$~GeV, a full
detector simulation study of signal and background 
processes including luminosity-weighted beam energy spectra and
beam-beam background event overlay is required.  Such an analysis
is ongoing, but results are not yet available.   For this report, we
extrapolate full simulation ILC results
 at $\sqrt{s}=500$~GeV\cite{Marchesini:2011aka}
down to $\sqrt{s}=250$~GeV in order to obtain the precision for
three parameter fits. Since one parameter fits were not
done  in the ILC studies at $\sqrt{s}=500$~GeV, we extrapolate
LEP~2 one parameter fit results at $\sqrt{s}\approx 200$~GeV~\cite{Schael:2004tq} up to
$\sqrt{s}=250$~GeV
and use the minimum of this extrapolation and the three parameter result
as estimates for the one parameter fits.

When extrapolating TGC statistical errors from one energy to another at least two effects must be considered\cite{TGCKarl}.  Clearly a $1/\sqrt{\sigma L}$ statistical factor must be included where
$\sigma$ and $L$ are the cross-section and integrated luminosity, respectively, at a particular center of mass energy.   Furthermore, a
factor inversely proportional to the center of mass energy squared, $s$, must be used to account for the energy dependence of the $SU(2)\times U(1)$ diagram cancellation.
In total  a factor $k_{ex}$  is used to extrapolate TGC statistical error from energy $A$ to energy $B$:
\beq    
k_{ex} =  \left(\sigma_A L_A \over \sigma_B L_B\right)^{\half} \left( s_A \over s_B \right)  \  .
\eeq{tgcEnergyExtrapolation}
We assume that
systematic errors are scaled by the same factor $k_{ex}$ as  results are extrapolated from
one energy to another.

\begin{table}
\begin{center}
  \begin{tabular} {|l|c||c|c|c||c|c|c||}
    \hline
 &   &  \multicolumn{3}{|c||}{total error ($\times 10^{-4}$) } & \multicolumn{3}{c||}{correlation} \\
    \hline
    Exp & $N_{par}$ & $ g^Z_1$  & $\kappa_\gamma$ & $\lambda_\gamma$ & $g^Z_1\ \kappa_\gamma$ &  $g^Z_1\ \lambda_\gamma$  & $\kappa_\gamma\ \lambda_\gamma$  \\
    \hline
    LEP~2     & 3     &  $516$  & $618$  & $376$  & -0.17 & -0.62 & -0.15 \\
    ILC~250   & 3        & $4.4$ & $5.7$ & $4.2$ & 0.63 & 0.48 & 0.35 \\
    \hline
    LEP~2     & 1     & $300$ & $626$ & $292$ & -- & -- & -- \\
    LHC      & 1     & $319$ & $1077$ & $198$ & -- & -- & -- \\
    HL-LHC   & 1      & $19$ & $160$ & $4$ & -- & -- & -- \\
    ILC~250   & 1       & $3.7$ & $5.7$ & $3.7$ & -- & -- & -- \\
    \hline

\end{tabular}
  \caption{TGC precisions for LEP~2, Run1 at LHC, HL-LHC and the ILC at $\sqrt{s}=250$~GeV with 2000~fb$^{-1}$ luminosity (ILC~250). The LEP~2 result is from ALEPH~\cite{Schael:2004tq} at  $\sqrt{s}\approx 200$~ GeV with
    0.68~fb$^{-1}$.  The LHC result is from ATLAS\cite{Aad:2014mda} at $\sqrt{s}=7$~TeV with 4.6~fb$^{-1}$.  The HL-LHC estimate is from a 2013 overview of HL-LHC physics~\cite{HLLHCMoenig}.}
\label{table:ilcextrap}
\end{center}
\end{table}

\begin{figure}
\begin{center}
   \begin{subfigure}{.49\hsize}
      \includegraphics[width=\textwidth]{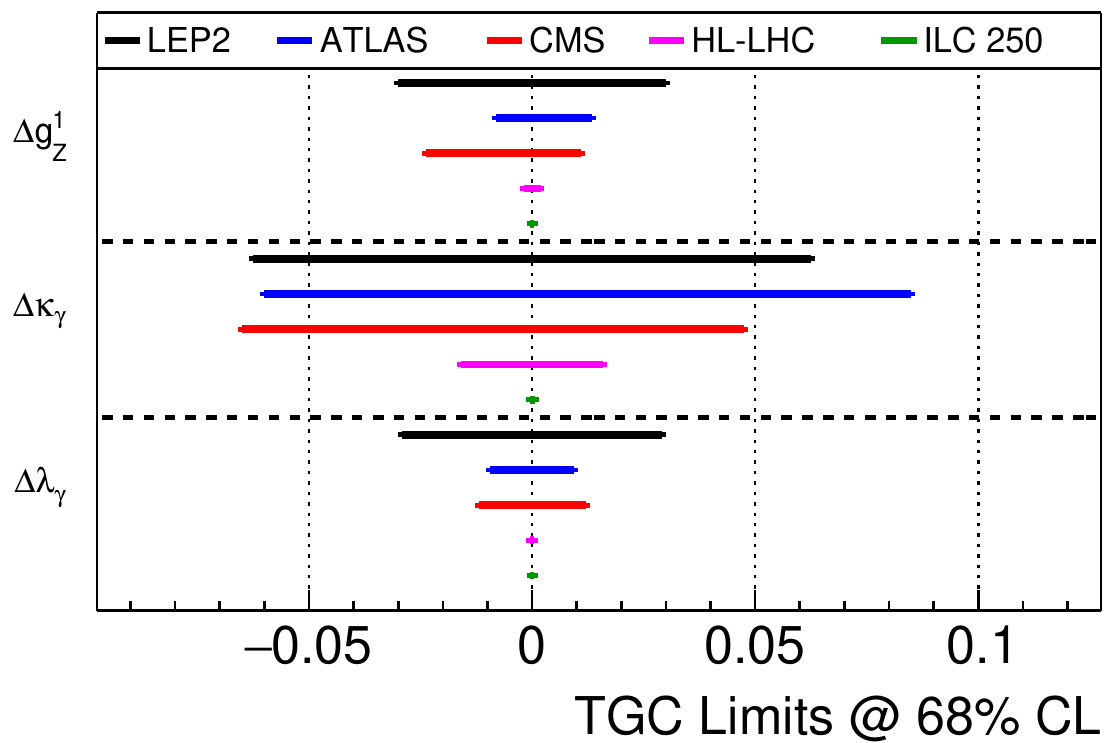}
         \subcaption{} \label{fig:TGCsingle}
   \end{subfigure} 
   \hspace{0.001\hsize} 
   \begin{subfigure}{.49\hsize}
      \includegraphics[width=\textwidth]{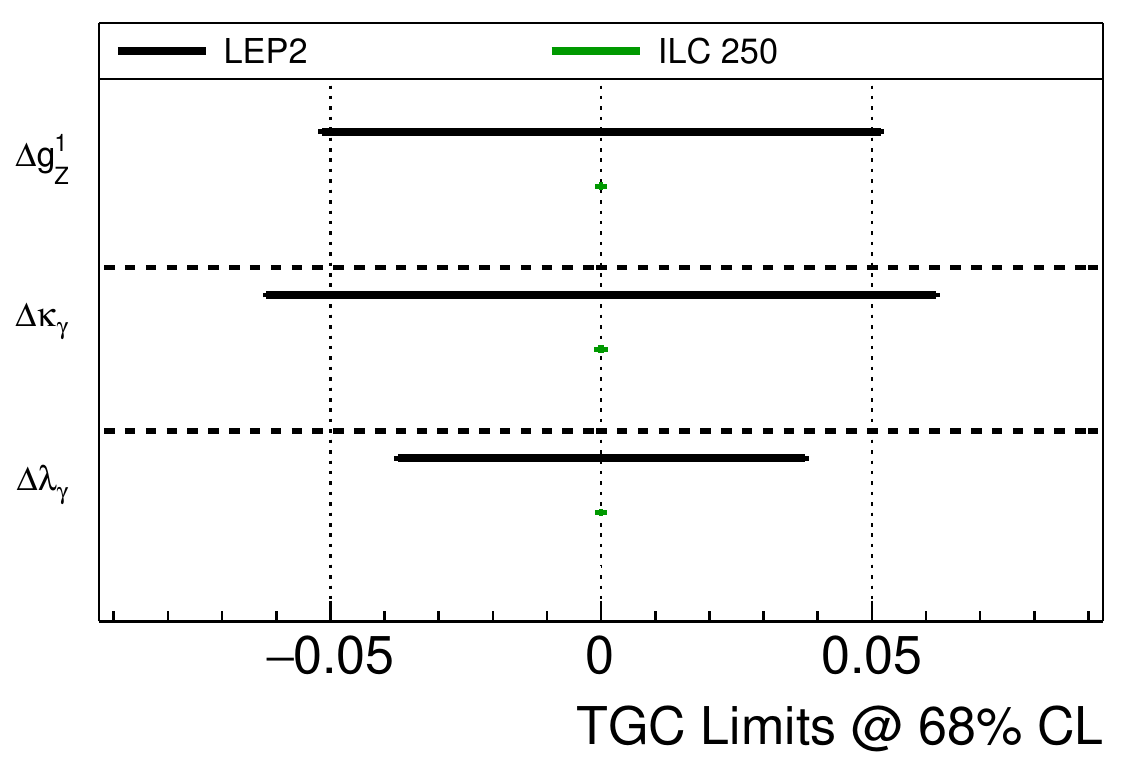}
         \subcaption{} \label{fig:TGCmulti}
   \end{subfigure}  
\end{center}
\caption{TGC precisions for LEP~2, Run1 at LHC, HL-LHC and the ILC at $\sqrt{s}=250$~GeV with 2000~fb$^{-1}$ luminosity (ILC~250)
using one parameter fits (a) and for LEP~2 and ILC~250 using three parameter fits (b).}
\label{fig:TGCplot}
\end{figure}

The TGC precisions for the ILC at $\sqrt{s}=250$~GeV with 2000~fb$^{-1}$ luminosity (ILC~250) are shown
in Table~\ref{table:ilcextrap} and Figure~\ref{fig:TGCplot}, along with results from LEP~2, LHC, and HL-LHC.
Results for one parameter fits where the other two 
anomalous couplings are set to zero are shown 
along with results for the full three parametrer fit.

At ILC~250 the three TGC's should be measured
  with accuracies ranging from  $4-6\times 10^{-4}$.
Comparing the one parameter ILC~250 
fit results with HL-LHC, the ILC~250 gives significantly better 
results for $g^Z_1$ and $\kappa_\gamma$ and
roughly the same result for $\lambda_\gamma$. 

The large sample of $W^+W^-$ and single-$W$ events at the ILC 250 also
offers an excellent setting for the measurement of the $W$ mass
through kinematic reconstruction of $W$ pair events 
and calorimetric comparison of
hadronic $W$ and  $Z$ decays.   These strategies are described in 
\cite{SnowmassEW,Wilson:2016tto}.   The systematics limit, which we
estimate as 2.4~MeV, should already be reached at the 250~GeV stage of
the ILC.

\section{Two-fermion production at 250~GeV}
\def\invfb{ \mbox{fb}^{-1} }
\newcommand{\ttbar}{ t \bar t}
\newcommand{\ffbar}{ f \bar f} 
\newcommand{\fonevZ}{F^{Z}_{1V}}
\newcommand{\fonevZh}{F^{Z}_{1V}}
\newcommand{\ftwovZ}{F^{Z}_{2V}}
\newcommand{\foneaZ}{F^{Z}_{1A}}
\newcommand{\ftwoaZ}{F^{Z}_{2A}}
\newcommand{\fonevA}{F^{\gamma}_{1V}}
\newcommand{\fonevAh}{F^{\gamma}_{1V}}
\newcommand{\ftwovA}{F^{\gamma}_{2V}}
\newcommand{\foneaA}{F^{\gamma}_{1A}}
\newcommand{\ftwoaA}{F^{\gamma}_{2A}}
\newcommand{\glZ}{g^{Z}_L}
\newcommand{\grZ}{g^{Z}_R}
\newcommand{\glA}{g^{\gamma}_L}
\newcommand{\grA}{g^{\gamma}_R}
\newcommand{\roots}{\sqrt{s}}

At an $\ee$ collider, the processes $\ee\to f\bar f$ can be measured with high precision for any SM fermion species.  In the $Z$ pole experiments at LEP and 
SLC, the measurement of two-fermion production in various final states gave
what are still the best measurements of the weak mixing 
angle $\sstw$~\cite{ALEPH:2005ab}.   At higher energies explored at LEP~2, interference of the $s$-channel photon and $Z$ diagrams produces order-1 forward-backward and polarization asymmetries. These can be used to probe for new effects, beyond the SM, that would be seen in interference with the SM contributions.   As for the physics topics presented in the previous two sections,  the ILC at 250 GeV  will lead to an improvement by more than an order of magnitude  in the sensitivity to these effects, due to the higher energy, the dramatically larger luminosity, and the use 
of beam polarization.

New physics contributions to $\ee\to f\bar f$ arise in a variety of models.   One possible source is a $Z'$ boson.   The LEP~2 experiments placed lower limits on the masses of  various types of $Z'$ bosons  in the range  500--800~GeV (and 1760~GeV for a sequential $Z$ boson)~\cite{Schael:2013ita}.   The corresponding 
limits from the 250~GeV ILC would be of order 5~TeV, 
comparable to the reach of LHC direct 
searches.  These limits would be improved by a factor 2 with ILC running at 
500~GeV.   The ILC searches are specific as to the flavor of the fermion species, the helicity of the coupling to electrons, and also, through the polarized forward-backward asymmetry, the helicity of the coupling to the final-state 
fermion.

Another possible source of corrections to the SM is the presence of 
extra dimensions, including the 
warped extra dimensions proposed in the model of Randall and Sundrum~\cite{Randall:1999ee} that also can be interpreted as a dual 
description of new strong interactions associated with the Higgs sector.  
Two-fermion processes, 
together with the Higgsstrahlung process~\cite{Angelescu:2017jyj}, are a very powerful probe for these models.  In these models, the new physics resonances
called Kaluza-Klein excitations modify the electroweak couplings to fermions
in a well-defined way.  For example, in the model proposed in 
\cite{Djouadi:2006rk},  only couplings to the (heavy) third generation quarks $(t,\,b)$ are modified. On the other hand, the model proposed in~\cite{Funatsu:2017nfm} predicts modifications to the couplings of all charged 
fermions. In both cases, one expects effects of the order of about 10\% already at a center-of-mass energy of 250~GeV.

An issue of particular interest is the measurement of the electroweak
 form factors of the $b$ quark.  The $b_L$ is certainly a heavy quark in the sense of the previous paragraph, since it is in the same $SU(2)\times U(1)$ multiplet as the top quark.  The $b_R$ might or might not be affected by Higgs strong interactions.   It is important to test for this possibility.  There are some 
tantalizing hints for non-standard behavior of the $b_R$.
There is a long-standing 3$\sigma$ discrepancy between the value of $\sstw^{\ell}$ derived from the $b$ forward-backward asymmetry at LEP and  the value obtained at the SLC using polarized beams~\cite{ALEPH:2005ab}.  Non-standard effects in the form factors of the $b_R$ might explain this 
discrepancy.  Hints for new physics are coming from Heavy Flavour Physics, as described, for example, in \cite{Neubert-Moriond2017}. In \cite{Megias:2017ove}, it is argued that the anomalies can be accommodated by requiring different degrees of compositeness of fermions in a dual theory.

\begin{figure}
\centering
\includegraphics[width=0.5\textwidth]{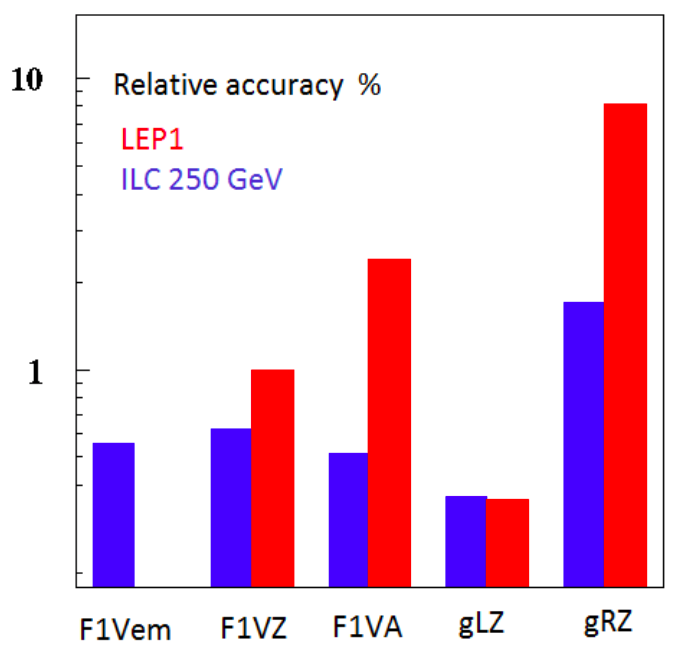}
\caption{\label{fig:manhattan-b} Comparisons of the precisions 
for the  electroweak form factors of the $b$ quark expected for the ILC at 250\,GeV for $500\,\invfb$ with those obtained by LEP. The figure is taken from~\cite{Bilokin:2017lco}}
\end{figure}

A recent full simulation study~\cite{Bilokin:2017lco} has investigated the process $\ee \rightarrow b\bar{b}$ at 250~GeV and for an  integrated luminosity of  0.5~ab$^{-1}$, shared between the different beam polarizations. Figure~\ref{fig:manhattan-b} shows that, already for this initial phase, the ILC precision for the $b$ quark form factors  will be much improved compared to the LEP results on the $Z$ pole, except for the case of the $b_L$ vector coupling, which is strongly constrained from  $BR(Z\to b\bar b)$.  A particularly interesting improvement is in the $b_R$ vector coupling $g_{RZ}$,  for which the ILC will outperform existing LEP results by about a factor of five. The measurements at the ILC will thus deliver the final word on the partial compositeness of the $b_R$, which is central to the open issues listed above.

  Based on \cite{Bilokin:2017lco} and on earlier studies of the $t$ quark~\cite{Amjad:2015mma}, it seems to be feasible to extend these quark form factor measurements to  the $c$ quark. This study can take advantage of the running at 250~GeV, since the decay products of the corresponding bottom or charm mesons are less boosted than at $\roots = 500$~GeV.  This is beneficial for the assignment of tracks to secondary vertices.

Fermion pair production is a powerful tool to set limits on fermion compositeness and may be probed by effective four-fermion vertices. The paper~\cite{AguilarSaavedra:2001rg} discusses the sensitivity of two-fermion production to new physics in terms of  these contact interactions at CM energies of $\roots = 500$~GeV and $\roots = 800$~GeV.  Extrapolating from this study,
we estimate that the ILC at 250~GeV will produce limits on the $\Lambda$ scale of contact interactions (interpreted as the inverse of the radius of a composite fermion) at roughly 60~TeV.

\section{Program of the ILC beyond 250~GeV}


We have seen above that the 250~GeV ILC has a great potential to discover BSM physics through precision measurements of the Higgs boson and various other electroweak processes, thanks to its well-defined initial state, its clean environment without QCD backgrounds, and its powerful polarized beams. 
We have also seen that these virtues of the ILC would allow the discovery of new particles already at 250~GeV. Such a new particle could include a dark matter particle, or a new particle that couples only very weakly to the SM particles, or new particles with a compressed mass spectrum that makes their detection extremely difficult at the LHC.

The real advantage of the linear collider is, however, its upgradability to higher energies by either expanding the length of the linacs or exploiting more advanced acceleration technology that would be available by the time of the upgrade. In Section~2, we outlined a reference staging scenario
 that consists of operation of the ILC at three energy stages: 250, 350, and 500 GeV. 
The physics goals of the higher energy stages have already
 been described in the reports \cite{Fujii:2015jha}  and \cite{Fujii:2017ekh}. 
 However, it is worth briefly recalling the main points here.

\begin{itemize}

\item 
The 350 GeV stage of the ILC will  enable us to carry out an energy scan of the  $t\bar{t}$ threshold.  This set of measurements will allow us to determine the threshold value of the top quark mass $m_t(1S)$  to 50 MeV. (See Section 3.2 of \cite{Fujii:2017ekh}.)  This is not only an improvement in accuracy over the expectation for $m_t$ at  the LHC, but also it is a measurement of a different quantity that is better defined theoretically and more closely connected to the top quark mass relevant for weak decay processes and grand unification.  The threshold top quark mass is closely related to the $\msb$ top quark mass; the conversion error is negligible if anticipated improvements in the value of $\alpha_s(m_Z)$ are realized.  If no deviations from the SM predictions are seen in other processes, this measurement will definitively settle the issue of the vacuum stability of the SM \cite{Buttazzo:2013uya}.

\item The 500 GeV stage of the ILC will provide a further improvement in the precision of the Higgs boson couplings accessible through Higgs decay by almost a factor 2 beyond the already strong results at 250 GeV.   We have already
demonstrated this in Table~\ref{tab:higgscouplings} and shown the implications for new physics discovery in Fig.~\ref{fig:chisq}.

\item  The 500 GeV stage of the ILC will give us access to two additional Higgs boson couplings that are not available at 250~GeV. The first of these is the Higgs coupling to $t\bar t$, which is measureable using the process $\ee\to t\bar t h$.  This Higgs boson coupling has large deviations from the SM expectation in many models in which the Higgs boson is composite or partially composite, for example~\cite{Malm:2014gha}.  The accuracy expected in this coupling with 500~GeV and 4~ab$^{-1}$ of data is 6\%.   The limited accuracy is due to the fact that 500~GeV is very close to the $t\bar t h$ threshold.  Comparable running at 550~GeV will enable an accuracy of 3\%.  At still higher energies,
 a 2\% determination is possible~\cite{SnowmassHiggs}

\item  The 500 GeV stage of the ILC will also bring us above threshold for the process $\ee\to Zhh$, from which it is possible to measure the triple Higgs coupling.  This measurement will give a first glimpse of the Higgs field potential beyond the measurement of the Higgs mass.   The measurement of the triple Higgs coupling is a crucial test for models of electroweak baryogenesis.  In models of this type, the Higgs phase transition must be of first order, and so a large deviation from the SM expectation for the potential is required~\cite{Noble:2007kk,Morrissey:2012db}.   The expected accuracy of the ILC measurement will be 27\%, sufficient to test this prediction.

Analyses of the triple Higgs coupling measurement typically assume that the only non-Standard effect is the change in the triple Higgs coupling and ignore the other possible effects of new physics on the observables.  In \cite{Barklow:2017awn}, these effects are studied within the EFT formalism and found to be potentially very substantial.  It is shown there that high precision measurements on single-Higgs processes are required to unambiguously interpret measurements of double-Higgs production.  At the ILC at 500~GeV, it is shown that the systematic error on the triple Higgs coupling from other new physics effects is smaller than 5\%, due to the high precision constraints that the ILC will give on the other 
16 relevant EFT coefficients.   There is no comparable strategy to address this point at hadron colliders.  In $pp$ collisions, many more EFT coefficients come into play, the constraints on these coefficients are weaker, and the 
dependence of the double Higgs production cross section on these coefficients is much stronger. 

\item The 500~GeV stage of the ILC will measure the form factors for the top quark couplings to the photon and $Z$ individually to accuracies below 1\%.
Models of composite Higgs bosons usually also entail partially composite top quarks.   This leads to substantial deviations from the SM expectations for the $Zt\bar t$ form factors, with characteristic differences between the couplings to $t_L$ and $t_R$ depending on the model.  A compilation of model predictions is given in \cite{Fujii:2015jha}.   These measurements give an additional, independent, route to the discovery of new physics associated with new strong interactions in the Higgs sector.

\item The 500~GeV stage of the ILC will improve the reach of searches for dark matter pair production, Higgsino production, and production of other  challenging proposed particles beyond the expectations given in Section 7.  The variety of new particles that can be discovered in direct production at 500~GeV is reviewed in \cite{Fujii:2017ekh}.

\item The 500~GeV stage of the ILC will substantially improve the discovery potential of the precision measurements of $\ee\to W^+W^-$ and $\ee\to f\bar f$ described in Sections 8 and 9.   The reach in terms of new physics scales will increase by almost  factor of 2. 

\end{itemize}

We do not know the ultimate energy reach of the ILC technology.  The ILC TDR documents a possible extention to 1~TeV based on current superconducting RF technology~\cite{Adolphsen:2013jya,Adolphsen:2013kya}.  However, the capabilities of superconducting RF accelerator are improving at a rapid pace; see, for example, \cite{Grassellino:2017bod}.  Over a longer term, we can imagine the development of advanced high-gradient accelerator  technologies that could enable an $\ee$ collider  at 10~TeV or higher in the ILC tunnel~\cite{Delahaye:2014kqa}.  If the 250~GeV ILC can discover the existence of new physics, later stages of the ILC Laboratory could explore this physics at its own natural energy scale.  The 250~GeV ILC is not an endpoint; rather, it is the first step toward a new method for uncovering physics beyond the SM.

\section{Conclusions}

The physics capabilities of the ILC at 250~GeV are formidable. 

  As we
have explained in this paper, this facility will provide
high-precision measurements of the couplings of the Higgs boson.
These coupling determinations will be model-independent and the values
of the output couplings will be absolutely normalized.  Neither
feature is possible at the LHC.   The precisions available at the
250~GeV stage of the ILC are close to 1\% for the Higgs coupling to the $b$
quark and below 1\% for the Higgs couplings to the $W$ and $Z$.  We
have demonstrated that this capability allows the discovery of new
physics for a variety of interesting models for which the predicted
particles are too heavy to be discovered at the LHC.

The ILC at 250~GeV also allows deep searches for exotic decays of the
Higgs boson.   Such decays are expected, in particular, in models in which
dark matter is a part of a ``hidden sector'' with no couplings to
Standard Model gauge bosons.  This program
of searching for dark matter using the Higgs is orthogonal to searches
for hidden sector particles with fixed target beams, a subject of much
recent interest, and it is no less important.

The ILC at 250~GeV will also carry out searches for pair production of
dark matter particles and other particles with small energy deposition
that are difficult to uncover at the LHC.  Although the energy
increase from LEP~2 is small, the integrated luminosity of the ILC
will be larger by a factor of 1000, leading to greatly improved reach
for many searches.   This luminosity increase and improvements in
detector technology will also allow us to greatly improve the
precision of  measurements in $\ee\to W^+W^-$ and $\ee\to f\bar f$ and
perhaps to expose new physics in those processes.

All of these approaches, and more, will benefit from operation of  the
ILC at higher energies.   The ILC at 250~GeV,
beyond the power of its own experiments, will be the first step along 
that road.

It is urgent today in particle physics to uncover physics beyond the
Standard Model by any route.   The experiments discussed in this
report give a number of strategies for searches for new physics that
are distinct from those currently being pursued at the LHC and
elsewhere.  These strategies have great potential.  But to exploit
them,   we must construct the next  $\ee$ collider.   The particle
physics community should make it a priority to fund and construct this
machine as quickly as possible.

\Acknowledgements

We are grateful to many people with whom we have discussed this
document.    Special thanks go to Rick Gupta, Howard Haber, 
 JoAnne Hewett, Ahmed Ismail, Hugh
Montgomery,  Francois Richard, Sabine Riemann, Heidi Rzehak, and
Graham Wilson.
   We are grateful for financial
support for our work from many agencies around the world.
TB  and MEP
were supported by the US Department of Energy under   
 contract DE--AC02--76SF00515.  TB, MB, CG, MH, RK, JL, and JR  are
 supported
 by the Deutsche Forschungsgemeinschaft (DFG)
 through 
the Collaborative Research
 Centre SFB 676 “Particles, Strings and the Early Universe”, projects
 B1 and B11.  CG is also supported by the European Commission through
the
Marie Curie Career Integration Grant 631962 and by the Helmholtz
Association
 through its  recruitment initiative.  HK and SJ are supported by
the  National Research
Foundation of Korea under grant 2015R1A4A1042542.  KF and TO are
 supported
by the Japan Society for the Promotion of Science (JSPS) under
Grants-in-Aid for Science Research 16H02173 and 16H02176.  JT is
supported by the  JSPS under Grant-in-Aid 15H02083.  MP is supported
by the U.S. National Science Foundation through grant PHY-1719877.  RP
is supported by the Quarks \& Leptons 
programme of the French IN2P3.

\appendix

\section{Projected ILC physics measurement uncertainties}

In Table~\ref{tab:results}, we summarize the projections for the
uncertainties in the measurements discussed in this report.

\begin{table}[p]
 \begin{center}
\begin{tabular}{lc|c|c|l}
Topic          &  Parameter   &250 GeV & 250 + 500 GeV &  units

\\  \hline 
Higgs          &   $ m_h $      &   14   & 14  &  MeV 
\\
                    &     $   g(hb\bar b)   $ & 
                  1.1 &  0.58   &    \% \\  
                  &    $   g(h c\bar c)    $   &   
                1.9 &   1.2 & \% \\
                  &    $    g(h g g)   $ & 
                  1.7 & 0.95     &      \% \\
                   &     $   g(hWW)     $ &  
                   0.67 &  0.34       &    \% \\
                &  $g(h\tau\tau)$ &   1.2  & 0.74&   \% \\
                    &   $   g(hZZ)   $    
                    &   0.68   &  0.35  &   \% \\
                
                 &    $    g(h \gamma \gamma)   $  &  1.2  &   1.0 &\% \\
                &    $   g(h\mu\mu)    $   &     5.6    &  5.1    &  \% \\
                   &    $    g(h \gamma Z)   $  &  6.6  &   2.6 &\% \\                   
                 &    $   g(h t\bar t)    $  &  - &   6.3     & \%,
               \\
                   &      $ g(hhh)     $         & - &  27     & \% \\
                    & $ \Gamma_{tot}$  & 2.5 &
                 1.6  & \% \\
                   & $ \Gamma_{invis}$ & 0.32&
                 0.29    & \%,   95\% CL \\
\hline
\hline
Top         &    $  m_t $        & - & 50  &  MeV ($m_t$(1S))  \\
        & $ \Gamma_t $  &-  & 60  &  MeV   \\ 
                &    $ g_L^\gamma   $ &-  &
               0.6  & \% \\
                          &    $ g_R^\gamma   $ &- &
               0.6  & \%\\
    &    $ g_L^Z   $ &- & 0.6  & \% \\
                &    $  g_R^Z   $ &- &
               1.0   & \% \\
                 &  $ \Re\ F_2^\gamma$  &- & 0.0014  & absolute \\
                 &  $\Re\  F_2^Z$  & -  & 0.0017  &absolute\\
                  &  $ \Im \ F_2^\gamma$  &- & 0.0014  & absolute \\
                 &  $ \Im\  F_2^Z$  & -  & 0.0020  &absolute\\
      \hline
$W$       &  $ m_W   $   & 2.4 MeV &   2.4  & MeV   \\
         &   $ g^Z_1     $    &        $   4.4\times 10^{-4}     $   &
        $ 1.1\times 10^{-4}  $ &   absolute \\
  &  $\kappa_{\gamma}   $   &   $ 5.7\times 10^{-4}     $    &    $  1.4 \times 10^{-4}  $   &
  absolute \\ 
   &  $\lambda_{\gamma}    $  &  $ 4.2 \times 10^{-4}    $    &    $  1.4 \times 10^{-4}   $&  absolute \\ 
\hline
Dark Matter     &  EFT $\Lambda$: D5  &  1.9 & 3.0 & TeV,
90\% CL \\ 
&  EFT $\Lambda$:  D8  &  1.8  & 2.8 & TeV,
90\% CL  \\ \hline
\end{tabular}
\caption{Projected accuracies of measurements of Standard Model
  parameters  for the 250 GeV stage of the ILC program and the
  complete program with 500 GeV running, as described in Section 2 
of this report.     The projected integrated  luminosities are:   2
ab$^{-1}$ at 250 GeV,  adding, for the full program,  0.2 ab$^{-1}$ at
350 GeV and 4 ab$^{-1}$ at 500 GeV.
 Initial state polarizations are as given at the end of Section 2.  Uncertainties are
  listed as $1\sigma$ errors (except where indicated),
  computed cumulatively at each stage of the program.  These estimated
  errors include
  both statistical uncertainties and theoretical and experimental systematic
  uncertainties. Except where indicated, errors in 
  percent (\%)  are fractional uncertainties
  relative to the Standard Model values. For dark matter, the
  effective field theory $\Lambda$ parameters are defined 
in \cite{Goodman:2010ku}.   More specific information for
each set  of measurements is given in  corresponding chapter of this report.}
\label{tab:results}
\end{center}
\end{table}

It is noteworthy that the improvements in the Higgs coupling analysis
reviewed in this paper allow us to claim stronger results for
precision coupling determinations than in our previous reports,
despite the fact that the proposed running energies are lower.  It is
especially interesting to compare the projections given in this report
with those given in our 2013 white paper 
\cite{SnowmassHiggs} and provided to the P5 panel
that formulated the US strategic plan for particle physics in
2014~\cite{Pfive}.
This comparison is shown in Table~\ref{tab:Snowmasstonow}.    All
entries refer to ``model-independent'' coupling determinations.
However, since the earlier analyses were done in the $\kappa$
formalism, their results are less model-independent than those presented in
this report.   The program that
 we have presented here, even for the first stage at
250~GeV, will fulfill the promises that we made in 2013.

\begin{table}
 \begin{center}
\begin{tabular}{l|cc|cc|r}
   & Snowmass 2013 :  & & this report : & \\ 
   Parameter   &ILC(500) & ILC(LumUp) &  250~GeV &  250+500~GeV & units
\\  \hline 
                   $   g(hb\bar b)   $ &   1.6 &  0.7 &
                  1.1 &  0.58   &    \% \\  
                   $   g(h c\bar c)    $   &   2.8 &  1.0 &
                1.9 &   1.2 & \% \\
              $    g(h g g)   $ &   2.3 & 0.9 &
                  1.7 & 0.95     &      \% \\
                $   g(hWW)     $ &   1.1 &   0.6 &
                   0.67 &  0.34       &    \% \\
               $g(h\tau\tau)$ &  2.3 &  0.9 &  1.2  & 0.74&   \% \\
                   $   g(hZZ)   $    & 1.0 &  0.5 &
                    0.68   &  0.35  &   \% \\
           $   g(h t\bar t)    $  &  14 &  1.9 &  - &   6.3     & \% \\
                  $ \Gamma_{tot}$  & 4.9 &   2.3 &  2.5 &
                 1.6  & \% \\
\hline
\end{tabular}
\caption{Projected accuracies of measurements of Higgs couplings
  presented in this report,  compared to the projected accuracies
  presented in the ``ILC Higgs White Paper'' prepared for Snowmass 2013
  \cite{SnowmassHiggs}, Table 6.1.   The column ILC(500) refers to the
  baseline program presented in that report:  250 fb$^{-1}$ at 250 GeV
  plus 500~fb$^{-1}$ at 500~GeV.  The column ILC(LumUp) refers to the
  upgrade discussed in that report, with a total of  1.15~ab$^{-1}$ at
250~GeV, 1600~ab$^{-1}$  at 500~GeV, and  2.5~ab$^{-1}$ at 1000~GeV.}
\label{tab:Snowmasstonow}
\end{center}
\end{table}

\end{document}